\documentclass[preprint,showpacs,titlepage,aps,prd,tightenlines,amsmath,byrevtex,nofootinbib]{revtex4}
\usepackage{graphicx}

\def\lsim{\raise0.3ex\hbox{$\;<$\kern-0.75em\raise-1.1ex
\hbox{$\sim\;$}}}
\def\gsim{\raise0.3ex\hbox{$\;>$\kern-0.75em\raise-1.1ex
\hbox{$\sim\;$}}}

\begin{document}
\title{Probing Non-Standard Neutrino Interactions with \\ 
Neutrino Factories}

\author{N.~Cipriano Ribeiro$^{1}$}
\email{ncipriano@fis.puc-rio.br}
\author{H.~Minakata$^{2}$}
\email{minakata@phys.metro-u.ac.jp}
\author{H.~Nunokawa$^{1}$}
\email{nunokawa@fis.puc-rio.br} 
\author{S.~Uchinami$^{2}$}
\email{uchinami@phys.metro-u.ac.jp}
\author{R.~Zukanovich Funchal$^{3}$}
\email{zukanov@if.usp.br}
\affiliation{
$^1$Departamento de F\'{\i}sica, Pontif{\'\i}cia Universidade Cat{\'o}lica 
do Rio de Janeiro, C. P. 38071, 22452-970, Rio de Janeiro, Brazil  \\ 
$^2$Department of Physics, Tokyo Metropolitan University, Hachioji, 
Tokyo 192-0397, Japan \\
$^3$Instituto de F\'{\i}sica, Universidade de S\~ao Paulo, 
 C.\ P.\ 66.318, 05315-970 S\~ao Paulo, Brazil
}


\vglue 1.0cm

\begin{abstract}
  We discuss the sensitivity reach of a neutrino factory measurement
  to non-standard neutrino interactions (NSI), which may exist as a
  low-energy manifestation of physics beyond the Standard Model.  We
  use the muon appearance modes $\nu_{e} \rightarrow \nu_{\mu}$/$\bar
  \nu_{e} \rightarrow \bar \nu_{\mu}$ and consider two detectors, one at
  $L=3000$ km and the other at $L=7000$ km; The latter is nearly at
  the magic baseline which is known to have a great sensitivity to
  matter density determination.
  Assuming the effects of NSI at the production and the detection are 
  negligible, we discuss the sensitivities to NSI and the simultaneous
  determination of $\theta_{13}$ and $\delta$ by examining  the effects 
  in the neutrino propagation of various systems in which two NSI 
  parameters $\varepsilon_{\alpha \beta}$ are switched on.  The
  sensitivities to off-diagonal $\varepsilon$'s are found to be
  excellent up to small values of $\theta_{13}$.  At $\sin^2
  2\theta_{13}=10^{-4}$, for example, $|\varepsilon_{e \tau}| \simeq
  \text{a few} \times10^{-3}$ at 3$\sigma$ CL for 2 degrees of freedom,
  whereas the ones for the diagonal $\varepsilon$'s are also
  acceptable, $|\varepsilon_{e e}| (|\varepsilon_{\tau \tau}|) \simeq
  0.1 (0.2)$ at the same CL.
  We demonstrate that the two-detector setting is powerful enough to
  resolve the $\theta_{13}$-NSI confusion problem, a notorious one
  which is thought to be an obstacle in determining $\theta_{13}$ and
  $\delta$.  We believe that the results obtained in this paper open
  the door to the possibility of using neutrino factory as a discovery
  machine for NSI while keeping its primary function of performing 
  precision  measurements of the lepton mixing parameters.
\end{abstract}


\maketitle

\section{Introduction}
\label{sec:intro}

In the last 10 years various neutrino experiments made it clear that
neutrinos have masses and lepton flavors mix \cite{MNS,review}.
Though the measurement of the values of $\theta_{13}$ and the CP phase
$\delta$ as well as determining the neutrino mass hierarchy still
elude us, there exist an array of experimental plans and ideas to make
the goal.  If successful they will bring us a more or less complete
picture of how neutrinos organize their mass spectrum and how 
lepton flavors mix.  When it happens, then, we will really be entering
into the era of precision measurement of lepton mixing parameters.
An intriguing question is whether this will serve just for precision
measurements of more or less known quantities, or could help us to
discover new physics beyond the Standard Model which is amended to
incorporate neutrino masses and mixing.

It has been suggested since long time ago that neutrinos may have
non-standard interactions (NSI) in addition to those dictated by the
Standard Model of electroweak
interactions~\cite{wolfenstein,valle,guzzo,Roulet:1991sm}.  If there
exists an energy scale for new physics at the TeV range, it is conceivable
that it produces higher dimension operators which affect the way how
neutrinos interact with matter, the point raised in \cite{grossmann}.
It was argued that such NSI possessed by neutrinos need not be subject
to the stringent constraints that the charged leptons have to obey
\cite{berezhiani}.  The existing constraints on NSI for neutrinos are
worked out in \cite{davidson,lep}.  If we follow naive dimensional
counting with four Fermi operators the order of magnitude of NSI is
expected to be $(m_Z / M_{NSI})^2 \sim 10^{-2}$ ($\sim 10^{-4}$) for
the energy scale of new physics $M_{NSI}$ of $\sim1$ (10) TeV
\cite{concha1}.
If we want to probe into such a tiny effect of NSI in the neutrino 
sector, extremely high sensitivities are required for experiments to 
detect deviations from the standard three flavor oscillations.
A large number of papers in the literature is devoted to discuss NSI
in various context including long-baseline accelerator experiments,
the atmospheric and the solar neutrino observations, and supernova
neutrinos~\cite{concha1,gago,confusion1,confusion2,ota2,ota1,concha2,
  fornengo,friedland,solar-NSI,sn-NSI,MINOS-NSI,hattori,honda,kopp,kopp2}.

It is the purpose of this paper to discuss the discovery potential of
NSI in a neutrino factory.
As is well known, a neutrino factory~\cite{nufact} is an ultimate
apparatus for precision measurements of the lepton mixing parameters.
Because of the capability of clean detection of muons including the
charge identification, its sensitivity to $\theta_{13}$ is expected to
go down to an extremely small value, somewhere in the range $\sin^2 2
\theta_{13} = 10^{-5} - 10^{-4}$~\cite{golden,nufactAPS,blondel}.
Then, it is natural to think about neutrino factory as a discovery
machine for NSI.

One of the key issues in the discussion of the discovery reach of NSI
is that we have to guarantee, at the same time, the ability of
accurately measure the lepton mixing parameters, especially 
$\theta_{13}$ and $\delta$, is left intact.
However, it is a difficult goal to achieve in particular when we go
down to very small values of $\theta_{13}$ and NSI contributions.  In
fact, it is known that the presence of NSI can confuse the measurement
of $\theta_{13}$ by mimicking its effect~\cite{confusion1,confusion2}.
Unless this problem is somehow solved, it is difficult to think about
a neutrino factory as a sensitive hunting tool for NSI and at the same
time as a viable apparatus for precision measurements of the lepton mixing
parameters.

We do overcome the confusion problem by our set-up.  We consider a two
detector setting with baselines at $L=3000$ km and at $L=7000$ km
\cite{confusion1}, which will sometimes be referred as the
intermediate and the far detectors, respectively.  Comparing the
yields taken by the two detectors is the key to resolve the confusion.
These features will be demonstrated in Secs.~\ref{zero-input} and
\ref{nonzero-input}.
We must note, however, that it is only by keeping the solar $\Delta
m^2$ terms that one can gain access to the CP phase $\delta$ which is
crucial to resolve the muddle between $\theta_{13}$ and NSI.
Therefore, we keep the solar $\Delta m^2$ terms throughout our
analytical and numerical treatment.

If the two detector setting is a special requirement for hunting NSI
it would not be easy to implement it into the project.  Fortunately,
it is {\em not} the case.  It has been
suggested~\cite{intrinsic,huber-winter} that combining measurements by
detectors at two baselines, one at 3000-4000 km and the other at
$\sim$7000 km, is powerful enough to resolve the parameter degeneracy
\cite{intrinsic,MNjhep01,octant} associated with the measurement of the
lepton mixing parameters, a notorious obstacle to its precision
determination.
The power of the two detector setting has been confirmed also by more
recent analysis \cite{huber_optim}.
 
In Sec.~\ref{NSI}, we present the general framework for probing NSI
through neutrino propagation in matter by switching on two NSI
contributions simultaneously and discuss some important points about
NSI effects in the $P(\nu_e \to \nu_\mu)$/$P(\bar \nu_e \to \bar \nu_\mu)$ 
oscillation probabilities, the channels explored in this paper.  
In Sec.~\ref{synergy}, we elucidate the powerfulness of the strategy 
to look for NSI in a neutrino factory by combining the measurements of 
two identical detectors at different baselines.
The assumptions we make about the neutrino factory parameters and the
two detector setup as well as the analysis we perform are described in
Sec.~\ref{analysis}.  In Sec.~\ref{zero-input}, we discuss the maximal
sensitivity to the various combinations of NSI parameters and their
effects on the precision measurements of the oscillation parameters
$\delta$ and $\theta_{13}$.  In Sec.~\ref{nonzero-input} we discuss
the accuracy of the determination of the NSI parameters as well as
their impact on the measurement of $\delta$ and $\theta_{13}$ in the
case they are sufficiently large to be measured by the neutrino
factory experiment.  The discovery reach to NSI, $\delta$ and
$\theta_{13}$ is presented in Sec.~\ref{disc-reach}.
Sec.~\ref{conclusion} is devoted to our final concluding remarks. In
Appendix~\ref{derivation}, we give details on the derivation of the
appearance probability $P(\nu_e \to \nu_\mu)$ in the presence of two
NSI parameters.

\section{Non-Standard Interactions of Neutrinos} 
\label{NSI}
\subsection{General features}
\label{general}

We consider NSI involving neutrinos of the type 
\begin{eqnarray}
{\cal L}_{\text{eff}}^{\text{NSI}} =
-2\sqrt{2}\, \varepsilon_{\alpha\beta}^{fP} G_F
(\overline{\nu}_\alpha \gamma_\mu P_L \nu_\beta)\,
(\overline{f} \gamma^\mu P f),
\label{LNSI}
\end{eqnarray}
where $G_F$ is the Fermi constant, and 
$f$ stands for the index running over fermion species in the earth, 
$f = e, u, d$, in which we follow \cite{davidson} for notation.  
$P$ stands for a projection operator and is either
$P_L\equiv \frac{1}{2} (1-\gamma_5)$ or $P_R\equiv \frac{1}{2} (1+\gamma_5)$. 
We summarize here the bounds on $\varepsilon_{\alpha\beta}^{fP}$ which are 
obtained in~\cite{davidson,lep} for the readers convenience:
\begin{eqnarray}
\left[
\begin{array}{ccc}
-0.9 < \varepsilon_{ee} < 0.75 & |\varepsilon_{e \mu}| \lsim 3.8 \times 10^{-4} & |\varepsilon_{e \tau}| \lsim 0.25 \\
 & -0.05 < \varepsilon_{\mu \mu} < 0.08 &  | \varepsilon_{\mu \tau} | \lsim  0.25  \\
  &  & | \varepsilon_{\tau \tau} | \lsim 0.4  \\
\end{array}
\right],
\label{bound}
\end{eqnarray}
bounds from Davidson {\it et al.} (LEP) are at 90\% (95\%) CL.

In this paper, we consider the effect of NSI in neutrino propagation
in matter. It is known that NSI can affect production and detection
processes of neutrinos so that a complete treatment must involve also
the latter two effects.
However, we should note that the muon storage ring which we assume as
a source of neutrinos is special as it is one of the cleanest.  Muon
decay has been studied with great precision and room for NSI is
smallest among the sources \cite{kuno}.  Because construction of the
muon storage ring will require an intense muon source, we assume that
the already tight constraints on NSI by muon decay would become much
more stringent at that time.  Therefore, we believe that to neglect NSI
in the production of neutrinos in fact gives a fair approximation,
unless we go down to extremely small values of
$\varepsilon_{\alpha\beta}$.

With regard to NSI at the detection this may be more debatable.
However, we call the readers' attention to the fact that upon
construction of the neutrino factory the near detector sitting in front of
the storage ring will give a severe bound on NSI \cite{davidson}.  We
also remark that even before that era several low energy neutrino
experiments may be able to place equally severe constraints on NSI
\cite{barranco,scholberg,bueno}.  The bounds from them are placed on
the product of NSI at the source and the detection.  But if the
constraint on NSI by muon decays is strongly constrained, this will 
translate into a stringent bound on NSI at the detection.
Furthermore, since we compare two identical detectors, at the 
intermediate ($L=3000$ km) and the far ($L=7000$ km) location,  
the NSI effect at detectors tend to cancel. 
Therefore, our approximation of ignoring NSI at the detection points may 
give a reasonable first approximation. 

To discuss the effects of NSI during neutrino propagation in matter we
will use the effective coefficients $\varepsilon_{\alpha\beta}$ as it is
traditional in this field.  They are defined as
$\varepsilon_{\alpha\beta} \equiv \sum_{f,P} \frac{n_f}{n_e}
\varepsilon_{\alpha\beta}^{fP}$.
where $n_f$ is the number density of the fermion species $f$ in matter. 
Normalizing by the electron number density $n_e$ leads to a simple 
structure of the effective Hamiltonian which governs the neutrino 
propagation in matter. 
Approximately, the relation 
$\varepsilon_{\alpha\beta} \simeq \sum_{P}
\left(
\varepsilon_{\alpha\beta}^{eP}
+ 3 \, \varepsilon_{\alpha\beta}^{uP}
+ 3 \, \varepsilon_{\alpha\beta}^{dP}
\right)$  
holds because of a factor of $\simeq$3 larger number of 
$u$ and $d$ quarks than electrons in iso-singlet matter.

The time evolution of neutrinos in flavor basis with non-standard 
neutrino matter interactions can be generically written as 
\begin{eqnarray} 
i {d\over dt} \left( \begin{array}{c} 
                   \nu_e \\ \nu_\mu \\ \nu_\tau 
                   \end{array}  \right)
 = \frac{1}{2E} \left[ U \left( \begin{array}{ccc}
                   0   & 0          & 0   \\
                   0   & \Delta m^2_{21}  & 0  \\
                   0   & 0           &  \Delta m^2_{31}  
                   \end{array} \right) U^{\dagger} +  
                  a \left( \begin{array}{ccc}
            1 + \varepsilon_{ee}     & \varepsilon_{e\mu} & \varepsilon_{e\tau} \\
            \varepsilon_{e \mu }^*  & \varepsilon_{\mu\mu}  & \varepsilon_{\mu\tau} \\
            \varepsilon_{e \tau}^* & \varepsilon_{\mu \tau }^* & \varepsilon_{\tau\tau} 
                   \end{array} 
                   \right) \right] ~
\left( \begin{array}{c} 
                   \nu_e \\ \nu_\mu \\ \nu_\tau 
                   \end{array}  \right)
\label{general-evolution}
\end{eqnarray}
where $U$ is the Maki-Nakagawa-Sakata (MNS) \cite{MNS} matrix, and
$a\equiv 2 \sqrt 2 G_F n_e E$ \cite{wolfenstein} where $E$ is the
neutrino energy and $n_e$ denotes the electron number density along
the neutrino trajectory in the earth.  $\Delta m^2_{i j} \equiv
m^2_{i} - m^2_{j}$ with neutrino mass $m_{i}$ ($i=1-3$).
Eq.~(\ref{general-evolution}) defines the framework for discussing
neutrino propagation in matter with NSI.

\subsection{Physics of neutrino propagation in matter with NSI; 
Two $\varepsilon$ system}
\label{2epsilon}

In this paper, we will be dealing with the cases where the following
pairs of $\varepsilon$ parameters are present at the same time:
$\varepsilon_{e \tau} - \varepsilon_{e e}$, 
$\varepsilon_{e \tau} - \varepsilon_{\tau \tau}$, 
$\varepsilon_{\tau \tau} - \varepsilon_{e e}$, 
$\varepsilon_{e \mu} - \varepsilon_{ee}$, 
$\varepsilon_{e \mu} - \varepsilon_{\tau \tau} $, and 
$\varepsilon_{e\tau} -\varepsilon_{e \mu}$.  
The systems with many $\varepsilon$'s are
complicated enough and a step by step approach is needed to grasp the
whole picture.
In Appendix~\ref{derivation}, we will derive the exact expressions and
the approximate formulas for the appearance oscillation probabilities
$P(\nu_e \to \nu_{\mu})$ in the two systems $\varepsilon_{e \tau} -
\varepsilon_{e e}$ and $\varepsilon_{e \mu} - \varepsilon_{e e}$.
To obtain tractable formulas we use perturbation expansion in terms of
small parameters, $s_{13} \simeq \delta_{21} \simeq \varepsilon_{e
  \tau}$ (or $\varepsilon_{e \mu}$) which we assume to have comparable
sizes $\sim 10^{-2}$, where $\delta_{21}$ is defined in
Appendix~\ref{derivation}.  For simplicity, we collectively denote
their order of magnitude as $\epsilon$ under the hope that no
confusion arises with the NSI elements $\varepsilon_{\alpha \beta}$.
See \cite{concha1,confusion2,ota1,kopp2,yasuda} for different
treatment of the perturbation series.  We will observe that in leading
order the appearance oscillation probabilities $P(\nu_e \to
\nu_{\mu})$ are of order $\epsilon^2$.
Of course, it reduces to the Cervera {\it et al.} formula
\cite{golden} if $\varepsilon_{\alpha \beta}$ are switched off.

We mention some notable features of the leading order formulas of
$P(\nu_e \to \nu_{\mu})$ in (\ref{Penu-2nd_etau}) and
(\ref{Penu-2nd_emu}), and make comments for better understanding of
our results.  To discuss simultaneously the $\varepsilon_{e \tau} -
\varepsilon_{e e}$ and $\varepsilon_{e \mu} - \varepsilon_{e e}$
systems, we use the notation $\varepsilon_{\alpha \beta}$ for
($\alpha, \beta)$ is either $(e \mu)$ or $(e \tau)$.
Since an off-diagonal element of $\varepsilon$ can have CP violating 
phase we have a new source for CP violation in addition to $\delta$ 
in the MNS matrix, as emphasized in \cite{concha1} in a somewhat 
different framework. In this context we notice that 
$\varepsilon_{\alpha \beta}$ comes in via three different ways: 

\vspace{0.2cm}
\noindent 
(a) $\varepsilon_{\alpha \beta}$ comes into the formula with its phase, 
$\phi_{\alpha \beta}$, in the form $\delta + \phi_{\alpha \beta}$ 
together with the leptonic Kobayashi-Maskawa (KM) phase 
$\delta$ in the MNS matrix. 
\cite{ota1,kopp}.

\vspace{0.2cm}
\noindent 
(b) $\varepsilon_{\alpha \beta}$ comes in by itself, i.e., 
in the form of ${\rm Re}( \varepsilon_{\alpha \beta} )$ or 
${\rm Im}( \varepsilon_{\alpha \beta} )$. 

\vspace{0.2cm}
\noindent 
(c) $\varepsilon_{\alpha \beta}$ comes in by the absolute magnitude squared, 
$|\varepsilon_{\alpha \beta}|^2$ without phase.

\vspace{0.2cm}
\noindent 
Terms of the type (a) and (c) survive even when the solar $\Delta
m^2_{21}$ is turned off, whereas terms of the type (b) arise only when
they are accompanied by the solar $\Delta m^2_{21}$.

We suggest an intuitive understanding of the fact that two phases come
together when $\Delta m^2_{21}$ is switched off.  The system without
$\Delta m^2_{21}$ contains effectively only two generations of
neutrinos and on physics ground the CP violating phase should be
unique.  Therefore, the phase of $\varepsilon_{\alpha \beta}$ and the
KM phase $\delta$ must come together.  From this reasoning, we
suspect that this feature holds even in the exact formula.
Of course, the two phases start to play their individual roles when
the solar $\Delta m^2_{21}$ is turned on, as in the form of (b) above
and the second line of the Cervera {\it et al.} formula
(\ref{Penu-zeroeps}), as seen in (\ref{Penu-2nd_etau}) and
(\ref{Penu-2nd_emu}).

Due to the special way through which the phase of $\varepsilon_{\alpha
  \beta}$ enters into the oscillation probability in the absence of
the solar $\Delta m^2_{21}$, we expect that confusions occur in the
experimental settings for which ignoring the solar $\Delta m^2_{21}$
gives a fair approximation.
We expect two types of confusion to occur: 

\begin{itemize}

\item Two phase confusion: Since they appear in the special
  combination $\phi_{\alpha \beta} + \delta$, it is obvious that two
  phase confusion occurs; the data allows determination of only the
  sum.  The confusion cannot be resolved even if the anti-neutrino
  channel is combined.  It can be resolved for relatively large
  $|\varepsilon_{\alpha \beta}|$ because the terms with $\Delta
  m^2_{21}$ start to play more important role.
  These features are in fact observed for $\varepsilon_{e \mu}$ in
  \cite{kopp}.
  We note that the symmetry under simultaneous shift of $\phi_{\alpha
    \beta}$ and $\delta$ by the same amount but with opposite signs is
  a symmetry of the whole system at the magic baseline.  We will see
  that this confusion shows up in our analysis.  (See
  Secs.~\ref{zero-input} and \ref{nonzero-input}.)

\item Phase-magnitude confusion: We expect that another type of
  confusion exists.  The term which contains the phases in
  (\ref{Penu-2nd_etau}) and (\ref{Penu-2nd_emu}) take the form
\begin{eqnarray} 
A |\varepsilon_{\alpha \beta}| \cos ( \phi_{\alpha \beta} + \delta ) + 
B |\varepsilon_{\alpha \beta}| \sin ( \phi_{\alpha \beta} + \delta ) = 
|\varepsilon_{\alpha \beta}| 
\sqrt{ A^2 + B^2 } \cos ( \phi_{\alpha \beta} + \delta - \xi) \, ,
\label{mag-phase}
\end{eqnarray}
where $\tan \xi = B/A$.  Therefore, varying the magnitude of the NSI
element can be compensated by adjusting the phase of
$\varepsilon_{\alpha \beta}$.
Unlike the two phase confusion, this confusion can in principle be
resolved if (1) $| \varepsilon_{\alpha \beta} |^2$ terms are important enough
to resolve the confusion, and/or (2) the anti-neutrino channel is combined, 
because the coefficients $A$ and $B$ are different between neutrino and 
anti-neutrino oscillation probabilities.
\end{itemize}

\section{Probing NSI by detectors at two baselines; 
their characteristics and synergy }
\label{synergy}

In this section, we indicate by qualitative level arguments that the 
$\nu_{\mu}$/$\bar \nu_{\mu}$ appearance measurement in neutrino factory at 
the magic baseline, 
\begin{eqnarray}
L = \frac{\sqrt{2} \pi} {G_F n_e}  = 7200 
\left(\frac{\rho}{4.5 \text{ g/cm}^3}\right)^{-1} \text{km}\, ,
\label{magicL}
\end{eqnarray}
and the synergy obtained when combined with measurement at 
$L \simeq 3000$ km provide a powerful method for probing 
non-standard neutrino-matter interactions. 
Powerfulness of the measurement at the magic baseline may be natural to 
expect for $\varepsilon_{ee}$ because it is effectively equivalent to 
measuring the electron number density in earth matter for which the 
accuracy of determination is known to be excellent 
\cite{mina-uchi,gandhi-winter}. 
We show in this paper that this statement is even more true for flavor 
off-diagonal NSI. 
We rely on $\nu_e \to \nu_{\mu}$/$\bar \nu_e \to \bar \nu_{\mu}$
appearance channels, the so called golden channels \cite{golden},
because of the matured muon detection technology.

\begin{figure}[htbp]
\includegraphics[width=1.0\textwidth]{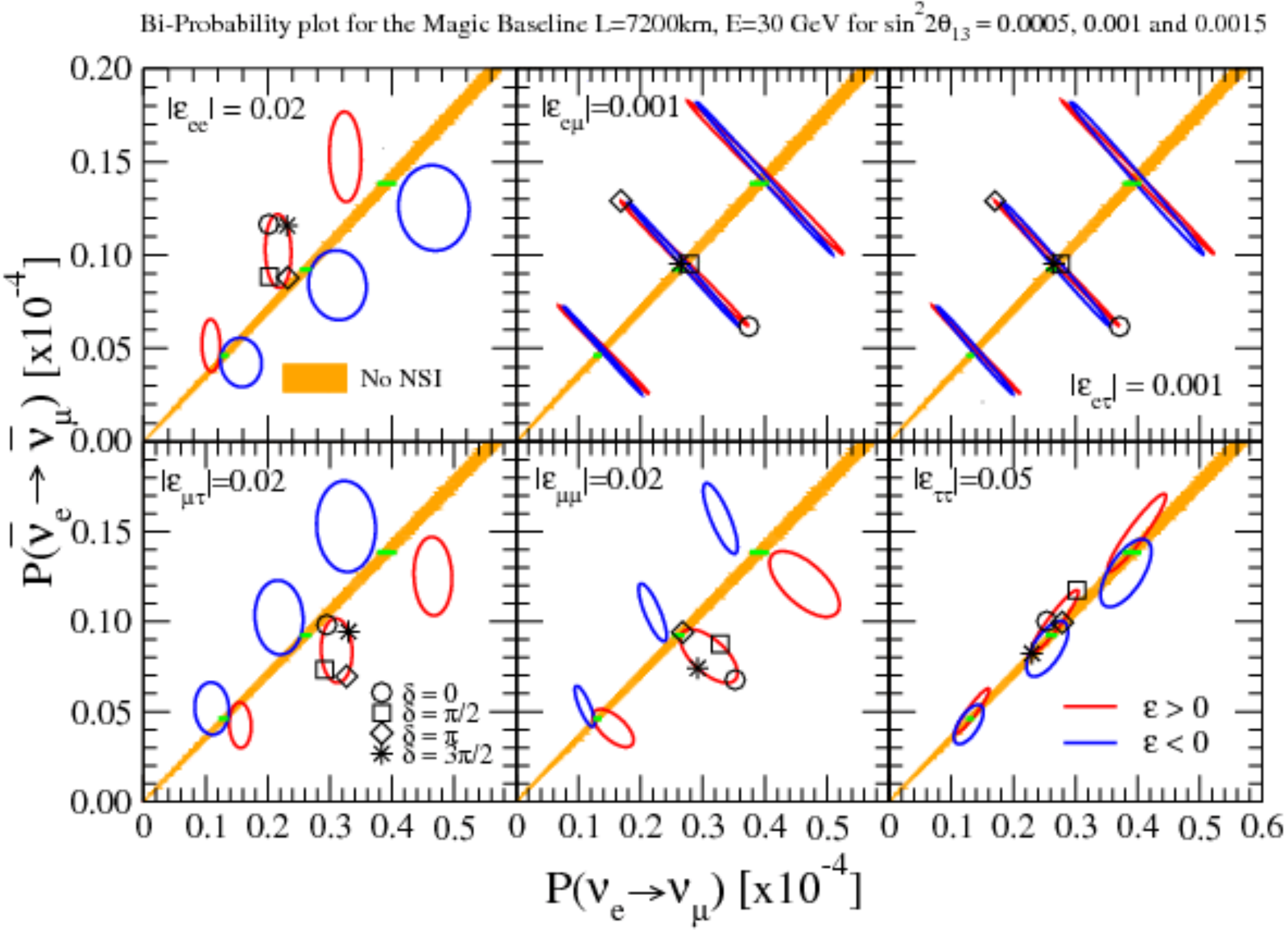}
\vglue -0.4cm
\caption{
Bi-probability plots in 
$P(\nu_e \to \nu_{\mu}) - P(\bar{\nu}_e \to \bar{\nu}_{\mu})$ space 
at the magic baseline, $L=7200$ km, for $E=$ 30 GeV, computed numerically 
using the constant matter density $\rho = 4.5$ g/cm$^3$ with 
the electron number density per nucleon equals to 0.5.  
The both axes is labeled in units of $10^{-4}$. 
In each panel only the indicated particular $\varepsilon_{\alpha \beta}$ 
is turned on. 
The upper (lower) panels, from left to right, correspond to the case
of non-vanishing $\varepsilon_{e e}$, $\varepsilon_{e \mu}$, and
$\varepsilon_{e \tau}$ ($\varepsilon_{\mu \tau}$, $\varepsilon_{\mu
\mu}$, $\varepsilon_{\tau \tau}$), respectively.  The red and the blue
ellipses are for positive and negative signs of $\varepsilon$,
respectively, for the cases with (from left to right) $\sin^2
2\theta_{13} = 0.0005$, 0.001, and 0.0015, as indicated in the
heading.
The values of the non-vanishing $\varepsilon$ are written in each
panel.  The orange colored region indicates the region spanned by
ellipses without NSI when $\theta_{13}$ is varied.  The green dots are
unresolved ellipses corresponding to the same values of $\sin^2
2\theta_{13}$ but without NSI.  The values of the standard lepton
mixing parameters are given in the caption of
Fig.~\ref{ee-et-tt-piby4}.  Only for the case of $\varepsilon >0$ and
$\sin^2 2\theta_{13} = 0.001$ we show the position corresponding to
the four different values of $\delta = 0, \pi/2, \pi$ and $3\pi/2$ by
the open circle, square, diamond and asterisk, respectively.  }
\label{bi-Pmagic}
\end{figure}

\subsection{Detector at the magic baseline as a sensitive probe to NSI}
\label{magic}

\begin{figure}[htbp]
\includegraphics[width=1.0\textwidth]{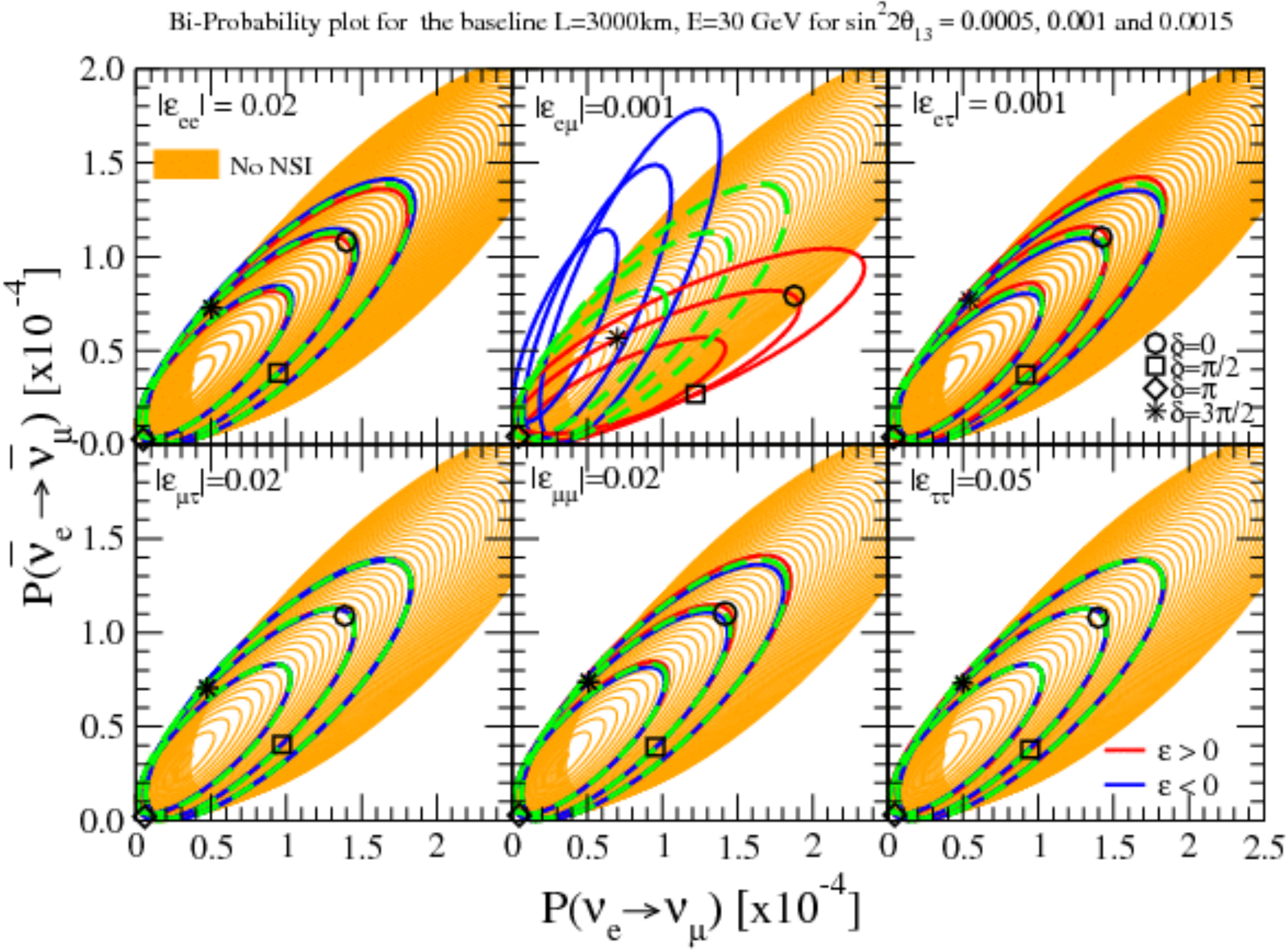}
\vglue -0.4cm
\caption{
The same as in Fig.~\ref{bi-Pmagic} but for the baseline $L=3000$ km
with the matter density $\rho = 3.6$ g/cm$^3$. 
The same values of $\varepsilon$ are used in each panel. 
In the left and right lower panels the ellipses with positive and 
negative sign of $\varepsilon$ overlap almost completely and 
each individual curve is not visible. 
The green ellipses which correspond to the same three values of 
$\sin^2 2\theta_{13}$ but without NSI are clearly visible. 
}
\label{bi-P3000}
\end{figure}

To illuminate that measurement at the magic baseline is a powerful
indicator for NSI, in particular for off-diagonal $\varepsilon_{\alpha \beta}$, we
present in Fig.~\ref{bi-Pmagic} the ellipses formed when $\delta$ is varied 
in the bi-probability space spanned by $P(\nu_e \to \nu_{\mu})$ and 
$P(\bar{\nu}_e \to \bar{\nu}_{\mu})$ \cite{MNjhep01} for various non-zero 
NSI parameters. The calculation was performed  numerically assuming the 
constant matter density $\rho = 4.5$ g/cm$^3$.  In each panel only the
indicated particular $\varepsilon_{\alpha \beta}$ is turned on.
Also shown in Fig.~\ref{bi-Pmagic} as an orange strip is the region covered by 
the ellipses when $\theta_{13}$ is varied for the cases without NSI. 
It approximately forms a narrow straight strip (or ``pencil'') because of 
vanishingly small effect of $\delta$ at the magic baseline \cite{BMW}. 
The remarkable feature of Fig.~\ref{bi-Pmagic} is that the effect of
NSI in the neutrino and the anti-neutrino probabilities is large for
electron-type off-diagonal terms, $\varepsilon_{e \tau}$ and
$\varepsilon_{e \mu}$, even though their size is extremely small,
$\varepsilon_{e \mu} = \varepsilon_{e \tau} = 10^{-3}$.  Notice that
the sizes of the other $\varepsilon$'s which give rise to effect of
similar magnitude as those of $\varepsilon_{e \mu}$ and
$\varepsilon_{e \tau}$ are larger by a factor of 20 ($\varepsilon_{\mu
  \tau}$ and $\varepsilon_{\mu \mu}$) and of $> 50 $
($\varepsilon_{\tau \tau}$).  This is the key point of our setting
which allows the extremely high sensitivity to $\varepsilon_{e \tau}$
and $\varepsilon_{e \mu}$.

There are some curious features in Fig.~\ref{bi-Pmagic}; The behavior
of the ellipses with $\varepsilon_{e \tau}$ and $\varepsilon_{e \mu}$
are distinct from the other cases having ellipses shrunk forming almost
lines.  At the same time it is also notable that they look almost
identical to each other.  Let us understand these characteristic
features of Fig.~\ref{bi-Pmagic}.

In Appendix~\ref{derivation} we derive the leading order formula for
$P(\nu_e \to \nu_{\mu})$ with $\varepsilon_{e \tau}$ or
$\varepsilon_{e \mu}$ as well as $\varepsilon_{e e}$ corresponding to
the ones derived by Cervera {\it et al.} \cite{golden} for the
standard case without $\varepsilon$.
At the magic baseline, $\frac{ a L}{4E} = \pi$, the formulas for the
$\nu_{\mu}$ appearance oscillation probability greatly simplifies.
With NSI represented by $\varepsilon_{e \tau}$ it is given by
\begin{eqnarray}
P(\nu_e \to \nu_{\mu}; \varepsilon_{e \tau}) 
&=&    
4 \frac{ ( \Delta m^2_{31} )^2 }{  (  a - \Delta m^2_{31} )^2  } 
s^2_{23} s^2_{13} 
\sin^2 \left(\frac{\Delta m^2_{31} L}{4E}  \right)  
\nonumber \\
&\hskip -2cm +& 
\hskip -1cm \frac{4 a  c_{23} s_{23}^2}{(a - \Delta m_{31}^2)^2}
  \Bigl[ 2  \Delta m_{31}^2 s_{13} \vert \varepsilon_{e\tau} \vert 
  \cos ( \delta + \phi_{e \tau} )
  + c_{23} a |\varepsilon_{e\tau}|^2 \Bigr] 
 \sin^2 \left( \frac{\Delta m_{31}^2 L}{4E} \right). 
\label{Pemu_magic-etau}
\end{eqnarray}
The corresponding formula for anti-neutrinos can be obtained by making
the replacement $a \rightarrow -a $, $\delta \rightarrow -\delta $,
and $ \phi_{e \tau} \rightarrow - \phi_{e \tau} $.
The formula with $\varepsilon_{e \mu}$ is very similar to
(\ref{Pemu_magic-etau}).  It is obtained by replacing $c_{23}
\varepsilon_{e \tau} $ by $ s_{23} \varepsilon_{e \mu} $ in the second
line of Eq.~(\ref{Pemu_magic-etau}).
Then, the value of $P(\nu_e \to \nu_{\mu})$ with $\varepsilon_{e \tau}$ 
and $\varepsilon_{e \mu}$  are numerically equal with the values of 
the parameters used in Fig.~\ref{bi-Pmagic}, 
the maximal value of $\theta_{23}$ and 
$| \varepsilon_{e \tau} | = | \varepsilon_{e \mu} |$. 
It explains the identical behavior of the probabilities 
with $\varepsilon_{e \mu}$ and $\varepsilon_{e \tau}$ seen in the second 
and the third upper panels  in Fig.~\ref{bi-Pmagic}. 
The ellipse in these panels shrink approximately to a line because
there is only $\cos ( \delta + \phi_{e \tau} )$ dependence in the
probability where $\phi_{e \tau}$ denotes the phase of $\varepsilon_{e
  \tau}$.  The form $\cos ( \delta + \phi_{e \tau} )$ means that the
shape and the location of shrunk ellipses in Fig.~\ref{bi-Pmagic} are
unchanged even when $\phi_{e \tau}$ is varied. The label of $\delta$
on the ellipse (if it is placed), of course, changes.

The length of the shrunk ellipse is equal to twice the coefficient of
$\cos ( \delta + \phi_{e \tau} )$ of the last term in
Eq.~(\ref{Pemu_magic-etau}).  With the values of the parameters used
(E=30 GeV, L = 7200 km, $\rho=4.5 \text{ g/cm}^3$, $\varepsilon_{e
  \tau}=0.001$, $\sin^2 2\theta_{13}=0.001$) it can be estimated as
$1.8 \times 10^{-5}$, which is in perfect agreement with the length
of the shrunk middle ellipse projected onto $P(\nu_e \to \nu_{\mu})$
axis in Fig.~\ref{bi-Pmagic}.
Therefore, we have understood the qualitative features of the flat 
ellipse at the magic baseline as well as its size.

We can also understand the reason why the effect of $\varepsilon_{\tau
  \tau}$ and $\varepsilon_{\mu \tau}$ are suppressed by deriving a
similar formula which contains one of them in an analogous way as in
Appendix~\ref{derivation}.  The leading order terms which involve one
of these $\varepsilon$'s are of order $\epsilon^3$ and hence
smaller.

\subsection{Detector at $L=3000$ km and the synergy expected when combined 
with the one at $L \simeq 7000$ km}
\label{3000km}

It is instructive to compare the similar plot for the intermediate detector 
at $L=3000$ km which is shown in Fig.~\ref{bi-P3000} to 
Fig.~\ref{bi-Pmagic} at $L=7200$ km.   
The same values of $\varepsilon$'s as in each corresponding panel 
in Fig.~\ref{bi-Pmagic} are used. 
The three representative values of $\sin^2 2\theta_{13}$ used are also 
the same as in Fig.~\ref{bi-Pmagic}. 
We immediately notice several clear differences. 
First of all, the effect of $\delta$ is large at $L=3000$ km for both
cases with and without NSI, as indicated by the blue and the red lines
(with NSI) and by the green ellipses as well as by wide span of the
ellipses indicated by the orange region (without NSI).  The ellipses
with NSI, except for the case with $\varepsilon_{e \mu}$,
are buried into this region. 
This is clearly the cause of the problem of confusion between 
$\theta_{13}$ and $\varepsilon$'s one encounters when one tries  
to measure $\theta_{13}$  and $\delta$ allowing for NSI; 
The system with $\varepsilon$'s can mimic the one with different values 
of $\theta_{13}$ but without NSI. 
In Fig.~\ref{bi-P3000} the difference between the confused parameters 
and the genuine ones are small, apart from the $\varepsilon_{e \mu}$ case, 
because we take small values of $\varepsilon$'s. 
It is also notable that the effect of the sign of the $\varepsilon$'s is 
not quite visible, which is nothing but a consequence of the two-phase 
degeneracy as explained in Sec.~\ref{2epsilon}. 
(See Sec.~\ref{degeneracy} for more about the degeneracy.)

When we try to determine the values of $\varepsilon$'s it is also a
bad news because there can be a severe confusion between NSI and the
standard oscillation effect with $\theta_{13}$ and $\delta$.  We will
see in Secs.~\ref{zero-input} and \ref{nonzero-input} that this
confusion for $\varepsilon$ determination is much severer than that in
$\theta_{13} - \delta$ determination at $L=3000$ km, which will be
manifested as a complicated island structure in the allowed region.
On the other hand, when we try to measure $\theta_{13}$ and $\delta$
at $L=3000$ km confusion due to the presence of $\varepsilon$'s is not
so significant {\em provided that the $\varepsilon$ is small} as we
see in Fig.~\ref{bi-P3000}.  In particular, as we will see, the
sensitivity to $\delta$ is good even after marginalization of
$\varepsilon$'s.  (See Fig.~\ref{nsi-th13-del-piby4-tau} -
\ref{nsi-th13-del-3piby2-mu} in Sec.~\ref{zero-input}.)

On the other hand, at $L \simeq 7000$ km the sensitivity to
$\varepsilon$'s is great though essentially there is no sensitivity to
$\delta$.  These consideration naturally suggest the possibility of
combining the intermediate and the far detectors to determine
simultaneously NSI parameters at the same time measuring $\theta_{13}$
and $\delta$.

The markedly different behavior of the bi-probability plots between
the $\varepsilon_{e \mu}$ and $\varepsilon_{e \tau}$ systems at $L
\simeq 3000$ km can be traced back to the difference between the third
terms in the analytic formulas (\ref{Penu-2nd_etau}) and
(\ref{Penu-2nd_emu}).
We should note that the almost identical behavior between the 
systems with $\varepsilon_{e \mu}$ and $\varepsilon_{e \tau}$ 
at $L=7000$ km, and the marked difference between them 
at $L \simeq 3000$ km makes it interesting by itself to compare their 
sensitivities at these two baselines and when they are combined. 
We will see later that sensitivity to $\varepsilon_{e \mu}$ is essentially 
determined by the effect at $L=3000$ km whereas that to 
$\varepsilon_{e \mu}$ is determined by the combination of
$L=3000$ km and 7000 km. 

\section{Analysis method}
\label{analysis}
\subsection{Assumptions}
\label{assumptions}

In our analysis in this paper, we make the following assumptions 
for the parameters of neutrino factory. We assume 
an intense muon storage ring which can deliver $10^{21}$ useful 
decaying muons per year. The muon energy is taken to be $50$ GeV. 
We assume 4 years running in neutrino and 4 years running in 
anti-neutrino modes, respectively. 
We assume the two magnetized iron detectors, one at baseline
$L=3000$ km and the other at $L=7000$ km which is close to but
not exactly the magic baseline $L_{\rm magic}$.\footnote{
By assuming the far detector at baseline somewhat off $L_{\rm magic}$,
we want to demonstrate that it is not quite necessary to place it
exactly at $L_{\rm magic}$.
First of all, we feel it unrealistic that one can place the far detector
at $L_{\rm magic}$ with a mathematical precision.
Apart from the problem of site availability, the exact magic
baseline cannot be determined prior measurement unless the
values of the relevant mixing parameters, the earth matter density
along the neutrino trajectory and its relationship with the effective
density for neutrino oscillation \cite{gandhi-winter} are precisely
known. Only a posteriori the matter density can be measured in situ in the
experiment\cite{mina-uchi}.
}
Each detector is assumed to have fiducial mass of 50 kton.
We consider the golden channels, $\nu_e \to \nu_{\mu}$ and $\bar{\nu}_e
\to \bar{\nu}_{\mu}$ in this paper.  For simplicity, we use the
constant matter density approximation throughout this work, and take
the Earth matter density along the neutrino trajectory as $\rho = 3.6
\text{ g/cm}^3$ and $\rho = 4.5 \text{ g/cm}^3$ for baselines $L=3000$
km and at $L=7000$ km, respectively.  The electron fraction $Y_e$ is
assumed to be 0.5.  We believe that using more realistic Earth matter
density profile will not change much our results.

In most part of our analysis we make the following simplifications: 
(1) We ignore all the background due to the misidentification
of muon charges and other causes.  
(2) We neglect the systematic uncertainties. 
Since the muon detection at high energies is supposed to be 
extremely clean in magnetized iron detectors the simplification (1) 
may not affect the results in a significant way apart from the case of 
extremely small $\theta_{13}$. 
With regard to the systematic errors we feel that no solid numbers 
are known yet in spite of the fact that great amount of efforts are made 
toward reliable estimation of them \cite{Cervera_nufact06}. 
Nevertheless, we will try to estimate to what extent the sensitivities 
we obtain in our analysis are affected by introduction of background 
and the systematic errors.  See Sec.~\ref{systematic}. 

We always take the normal mass hierarchy as an input. 
We will consider in this paper various cases in which only two
different flavor elements of NSI ($\varepsilon_{\alpha \beta}$) are
turned on at the same time.  We also assume that $\varepsilon_{\alpha
  \beta}$ is real, leaving the interesting topics of complex phase
effects to elsewhere. 

\subsection{Analysis procedure}
\label{procedure}

We define the $\chi ^2 $ function as follows,

\begin{equation}
\chi ^2 \equiv 
\min_{\theta_{13},\delta,\varepsilon}
\sum_{i=1}^3
\sum_{j=1}^2
\sum_{k=1}^2
\frac{ 
\left[ N^{\text{obs}}_{i,j,k} -
N^{\text{theo}}_{i,j,k}(\theta_{13},\delta,\varepsilon) 
\right]^2 }
{N^{\text{theo}}_{i,j,k}(\theta_{13},\delta,\varepsilon) },
\label{eq:chi2}
\end{equation}
where 
$N^{\text{obs}}_{i,j,k}$ is the number of observed (simulated) events
computed by using the given input parameters 
and $N^{\text{theo}}_{i,j,k}$ is the theoretically expected 
number of events to be varied in the fit by freely varying 
the mixing and NSI parameters. 
Since we ignore the systematic uncertainties, the denominator 
in (\ref{eq:chi2}) represent the statistical uncertainties. 
The summation with respect to indices $i,j$ and $k$ imply energy (3
bins), baseline (3000 km or 7000 km), and the type of neutrinos
(neutrino or anti-neutrino), respectively.  The intervals of 3 energy
bins we consider are 4-8 GeV, 8-20 GeV and 20-50 GeV for neutrinos and
4-15 GeV, 15-25 GeV and 25-50 GeV for anti-neutrinos. 

The theoretically expected number of events are computed as 
\begin{equation}
N^{\text{theo}}(\theta_{13},\delta,\varepsilon) 
= n_{\mu} T M \frac{10^9 N_{\text{A}}}{m^2_\mu \pi} 
\frac{E^2_\mu}{L^2} \int_{E_{\text{min}}}^{E_{\text{max}}} 
g(E) \, \sigma_{\nu_{\mu}(\bar{\nu}_{\mu})}(E) \, 
P_{\nu_e \to \nu_\mu (\bar{\nu}_e \to \bar{\nu}_\mu) } 
(E;\theta_{13},\delta,\varepsilon) dE,
\label{emu}
\end{equation}
where $n_{\mu}$ is the number of useful muon decays per year, $T$ is
the exposure period (in years), $M$ is the detector mass (in ktons),
$N_A$ is the Avogadro's number, $m_\mu$ is the muon mass, $E_{\mu}$ is
the energy of the stored muons, $L$ is the baseline,
$\sigma_{\nu_{\mu}(\bar{\nu}_{\mu})}(E)$ is the charged current
interaction cross section for $\nu_\mu$ and $\bar {\nu}_\mu$, and
$P_{\nu_e \to \nu_\mu (\bar{\nu}_e \to \bar{\nu}_\mu) }
(E;\theta_{13},\delta,\varepsilon)$ is the oscillation probability.
In this work, we considered the case where $E_\mu = 50$ GeV, $M = 50$
kton, $T$ = 4 yr for both neutrinos and anti-neutrinos, and $n_{\mu} =
10^{21}$ per year.
The function $g(E)$ which is given as
\begin{equation}
g(E) \equiv  \displaystyle 12\frac{E^2}{E_\mu^3}\left(1-\frac{E}{E_\mu}\right) 
\end{equation}
is the unoscillated $\nu_e$ or $\bar{\nu}_e$
energy spectrum normalized to 1.

We assume, for simplicity, the detection efficiency is 100\%.\footnote{
If the efficiency is $f \times$100\% we are effectively assuming the fiducial 
mass of the detector of $50/f$~kton. 
In the current estimate $f$ is expected to be about 0.8 apart from 
an extremely low energy region $\sim 5$~GeV 
\cite{Cervera_nufact06}. }
%
We neglect the finite energy resolution in the detectors. 
Since the number of energy bins are small (=3) inclusion 
of the energy resolution will not alter the results in a significant way. 

The observed number of events are computed exactly in the same way but
using the given input parameters of $\theta_{13}$, $\delta$ 
and NSI parameters ($\varepsilon$), so that $\chi^2_{\text{min}}= 0$ 
at the best fit point by construction. 

Using the $\chi^2$ function defined in Eq.~(\ref{eq:chi2}), we define the 
allowed (sensitivity) regions by the commonly used condition, 
$\Delta \chi^2 \equiv \chi^2-\chi^2_{\text{min}}$ = 2.3, 6.18 and 11.83 
for 1, 2 and 3 $\sigma$ confidence level (CL) for 2 degrees of freedom (DOF), 
unless otherwise stated.

\section{Sensitivity to Non-Standard Interactions and  
measurement of $\theta_{13}$ and $\delta$ with NSI}
\label{zero-input}

In this section, we discuss the sensitivity to NSI by the two detector
setting.  We also discuss the accuracy of the determination of the
mixing parameters $\theta_{13}$ and $\delta$ in the presence of NSI.
For these purposes we take the input values of two $\varepsilon$'s to
be zero (or equivalently vanishingly small) but freely vary them in
fitting the data.  The case with non-zero input values of
$\varepsilon$'s will be dealt with in Sec.~\ref{nonzero-input}.
To demonstrate the synergy between the intermediate and the far
detectors we present the sensitivity to NSI for each detector as well
as the combined one throughout this and the next sections.  As a
typical value of $\theta_{13}$ we consider the case of $\sin^2
2\theta_{13} = 10^{-3}$, though we also discuss the case with $\sin^2
2\theta_{13} = 10^{-4}$ to show how the sensitivities depend on
$\theta_{13}$.  We have examined the four values of $\delta$, $\pi/4$,
$\pi/2$, $\pi$, and $3\pi/2$ as representative cases.  However, we
present only part of the figures we have drawn not to make this paper
too long.

\subsection{Constraining NSI; Case of zero input}
\label{NSI-zero}

\begin{figure}[b]
\vglue -0.5cm
\begin{center}
\includegraphics[width=1.0\textwidth]{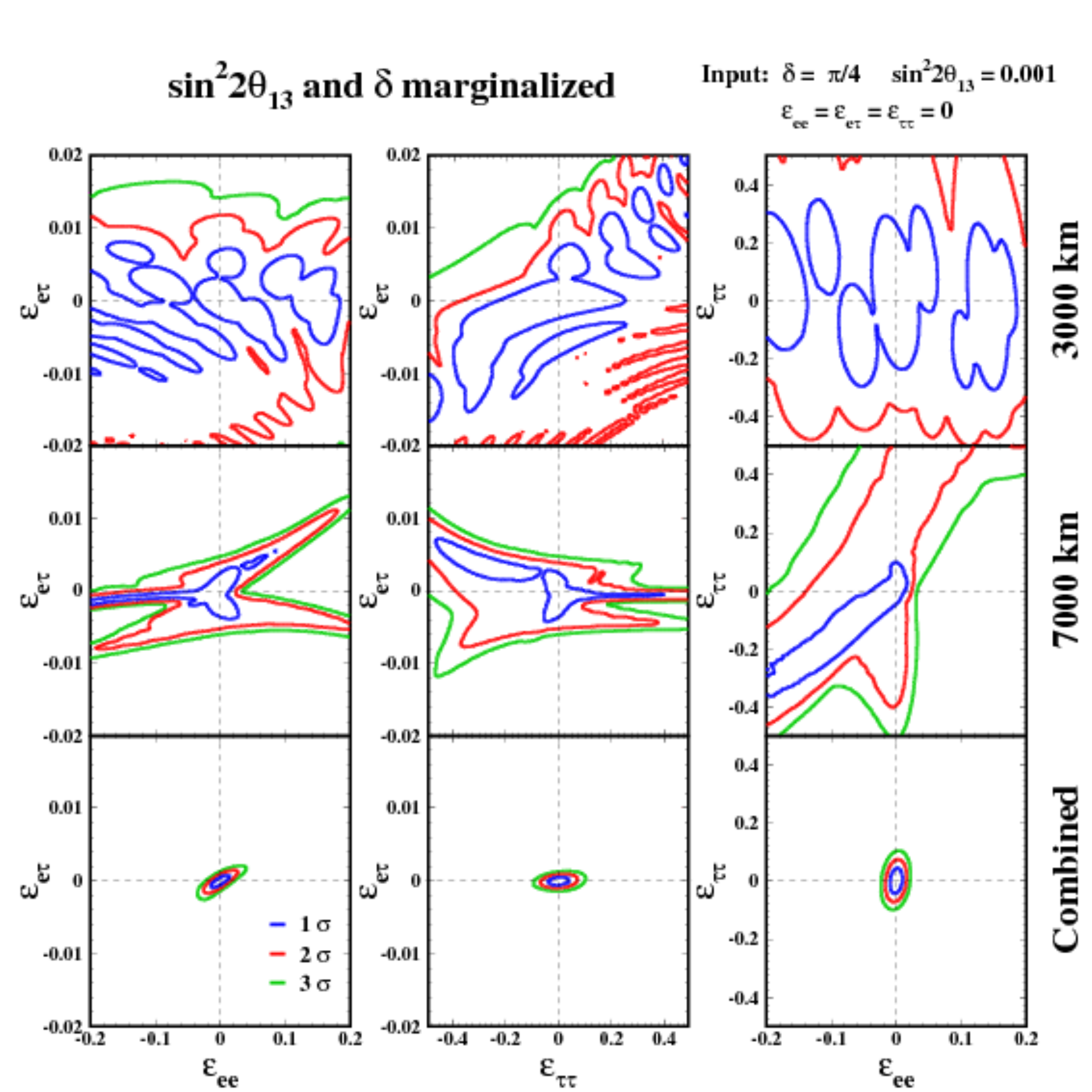}
\end{center}
\vglue -0.5cm
\caption{
Allowed regions projected into the plane of 
2 NSI parameters, $\varepsilon_{ee}$-$\varepsilon_{e\tau}$
(left panels), $\varepsilon_{\tau\tau}$-$\varepsilon_{e\tau}$ (middle panels)
and $\varepsilon_{ee}$-$\varepsilon_{\tau\tau}$ (right panels) 
corresponding to the case where 
the input parameters are $\sin^2 2\theta_{13} = 0.001$ 
and $\delta = \pi/4$ and no non-standard interactions
(or all the $\varepsilon$'s are zero), 
for $E_\mu$ = 50 GeV and 
the baseline of $L=3000$ km (upper panels), 7000 km (middle horizontal panels) 
and combination (lower panels). 
The thin dashed lines are to indicate the input values of 
$\varepsilon_{\alpha \beta}$.
The fit was performed by varying freely 4 parameters, 
$\theta_{13}$, $\delta$ and 2 $\varepsilon$'s 
with $\theta_{13}$ and $\delta$ being marginalized. 
The number of muons decays per year is $10^{21}$, 
the exposure considered is 4 (4) years for neutrino (anti-neutrino),
and each detector mass is assumed to be 50 kton. 
The number of energy bins considered is three.
The other standard oscillation parameters are fixed as
$\Delta m^2_{23} = 2.5 \times 10^{-3}$ eV$^2$, $\sin^2 \theta_{23} = 0.5$, 
$\Delta m^2_{12} = 8.0 \times 10^{-5}$ eV$^2$ and $\sin^2 \theta_{12} = 0.31$.
}
\label{ee-et-tt-piby4}
\end{figure}

\begin{figure}[htbp]
\vglue -0.5cm
\begin{center}
\includegraphics[width=1.0\textwidth]{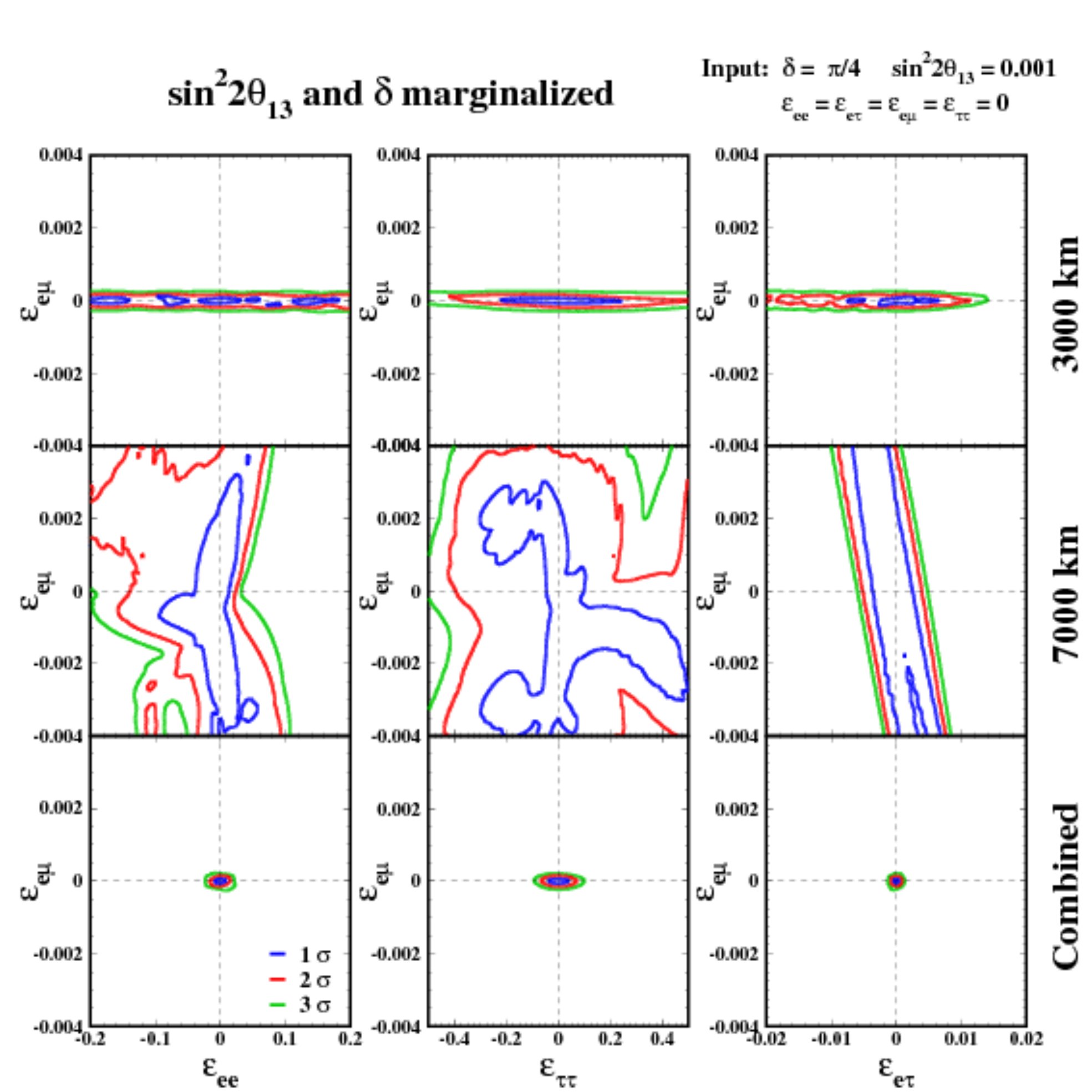}
\end{center}
\caption{The same as in Fig.~\ref{ee-et-tt-piby4} but  
for a different combination of 2 $\varepsilon$'s, 
$\varepsilon_{ee}$-$\varepsilon_{e\mu}$
(left panels), $\varepsilon_{\tau\tau}$-$\varepsilon_{e\mu}$ (middle panels)
and $\varepsilon_{e\mu}$-$\varepsilon_{e\tau}$ (right panels). 
}
\label{ee-em-mm-piby4}
\end{figure}

\begin{figure}[htbp]
\vglue -0.5cm
\begin{center}
\includegraphics[width=1.0\textwidth]{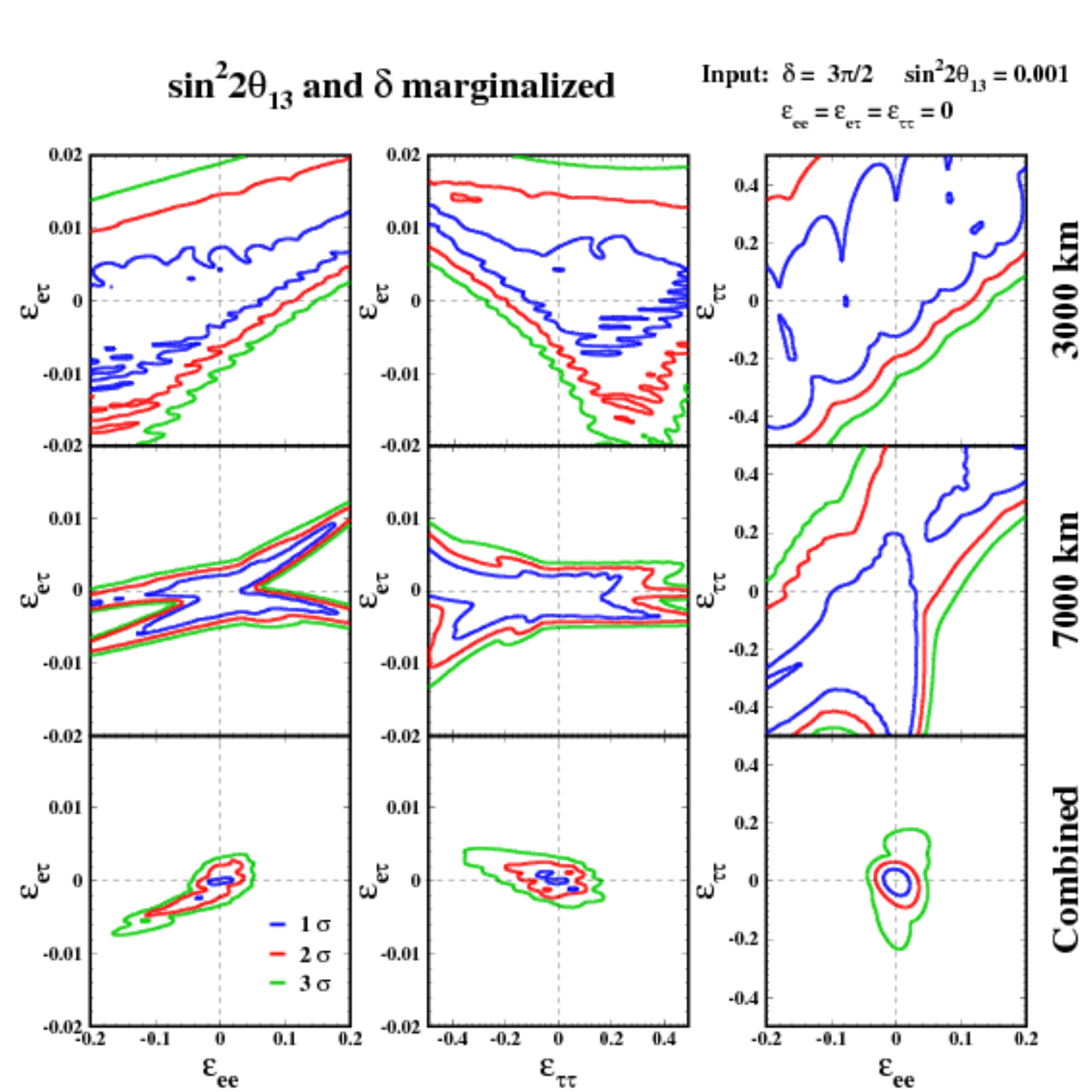}
\end{center}
\caption{The same as in Fig.~\ref{ee-et-tt-piby4} but with $\delta = 3\pi/2$.
}
\label{ee-et-tt-3piby2}
\end{figure}

\begin{figure}[htbp] 
\vglue -0.5cm
\begin{center}
\includegraphics[width=1.0\textwidth]{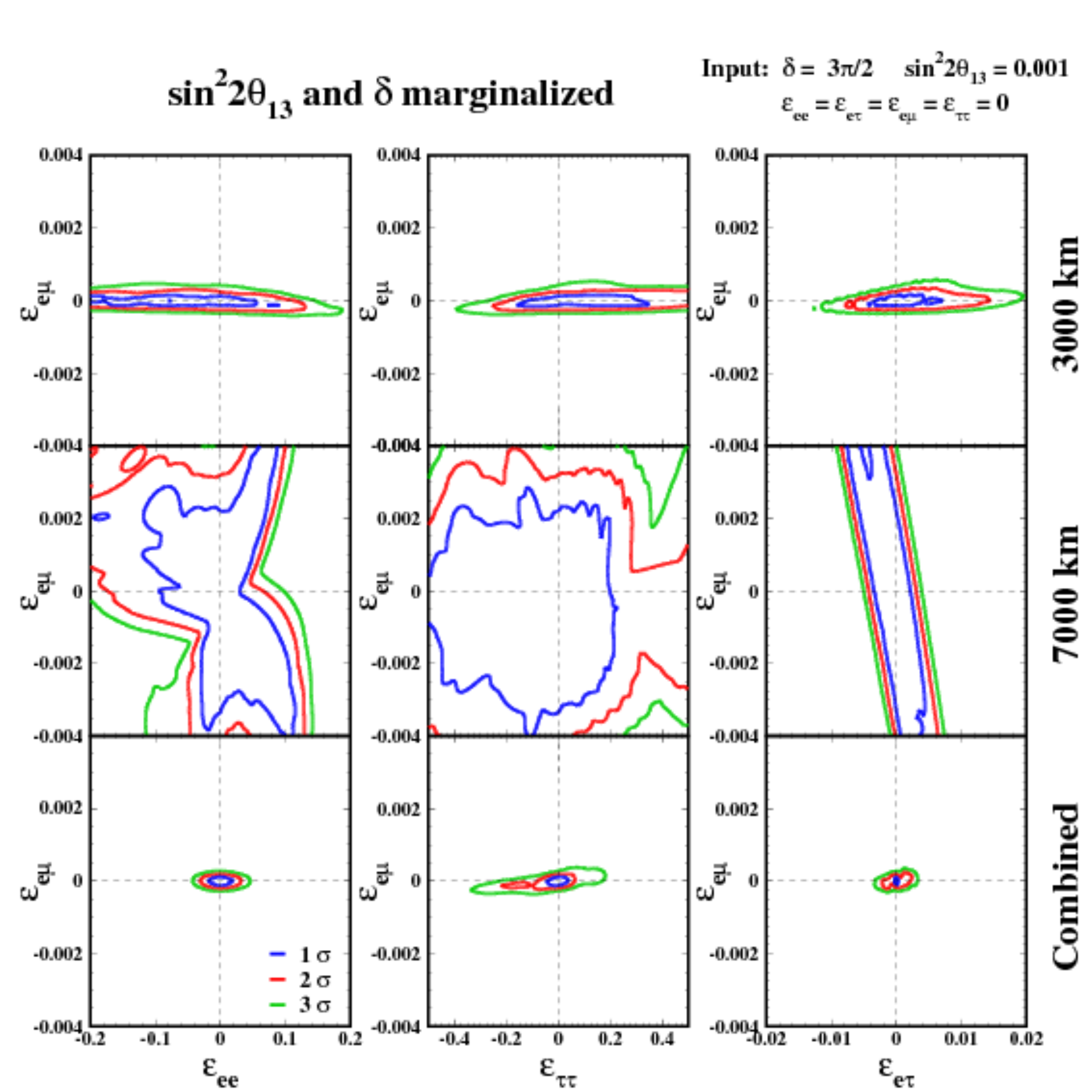}
\end{center}
\caption{The same as in Fig.~\ref{ee-em-mm-piby4} but with $\delta = 3\pi/2$.}
\label{ee-em-et-3piby2}
\end{figure}

In Figs.~\ref{ee-et-tt-piby4}-\ref{ee-em-et-3piby2}, we present the
constraint on NSI that can be imposed by the neutrino factory
measurement defined in Sec.~\ref{assumptions} with various selected
combination of two $\varepsilon$ parameters.\footnote{
It appears to us that ragged behavior of the contours seen in some 
of the plots is of physical origin due to the complicated structure 
of four-dimensional $\chi^2$ function. 
Some of the structures, however, could be smoothen to a certain 
extent by introducing a finer grid and the finite energy resolution. 
}
%
Fig.~\ref{ee-et-tt-piby4} and \ref{ee-em-mm-piby4} are for $\delta=\pi/4$ 
whereas 
Fig.~\ref{ee-et-tt-3piby2} and \ref{ee-em-et-3piby2} are for $\delta=3 \pi/2$. 
The left, the middle, and the right panels of 
Figs.~\ref{ee-et-tt-piby4} and \ref{ee-et-tt-3piby2} 
(Figs.~\ref{ee-em-mm-piby4} and \ref{ee-em-et-3piby2}) 
are for the combination, showing horizontal - vertical axes,  
$\varepsilon_{ee} - \varepsilon_{e \tau}$,  
$\varepsilon_{\tau \tau} - \varepsilon_{e \tau}$,  
$\varepsilon_{ee} - \varepsilon_{\tau \tau}$ 
($\varepsilon_{ee} - \varepsilon_{e \mu}$,  
$\varepsilon_{\tau \tau} - \varepsilon_{e \mu}$,  
$\varepsilon_{e \tau} - \varepsilon_{e \mu}$), respectively. 
The top, the middle, and the bottom panels are, respectively, for the
baselines $L=3000$ km, $L=7000$ km, and the two baselines combined.
The blue, the red, and the green curves are the allowed contours at 
1$\sigma$, 2$\sigma$ and 3$\sigma$ CL for 2 DOF, respectively.

We observe that from the results presented in
Fig.~\ref{ee-et-tt-piby4} through Fig.~\ref{ee-em-et-3piby2}, the
detector at $L=3000$ km alone does not have good resolution power 
for the possible existence of $\varepsilon$'s, except for 
$\varepsilon_{e\mu}$. The parameter $\varepsilon_{e \mu}$ is special
since its impact on the oscillation probability is so large (as seen
in Fig.~\ref{bi-P3000}) that the measurement at $L=3000$ km alone can
give a strong constraint on $\varepsilon_{e \mu}$ which seems
consistent with the result obtained in Ref.~\cite{kopp}.
Apart from the cases which involve $\varepsilon_{e \mu}$, 
the effect of the simultaneous presence of two $\varepsilon$'s is manifest 
in the appearance of many correlated regions/islands, though 
the precise shapes of the regions depend on which combination of 
$\varepsilon$'s is turned on and on which input value of $\delta$ is used. 

A similar statement applies to the case of the detector at L=7000 km. 
But, the correlation between $\varepsilon$'s is quasi one-dimensional 
for most of the combinations though we see branch structure in 
$\varepsilon_{ee} - \varepsilon_{e \tau}$ case and sizable width 
in the $\varepsilon_{\tau \tau} - \varepsilon_{e \tau}$ case. 
Overall, the constraints on the diagonal elements, 
$\varepsilon_{ee}$ and $\varepsilon_{\tau \tau}$, are much looser 
compared to those on $\varepsilon_{e \tau}$ and $\varepsilon_{e \mu}$. 
The latter feature comes from high sensitivities to the 
off-diagonal $\varepsilon$'s expected at the magic baseline as 
discussed in Sec.~\ref{magic}, while less sensitivity to the former 
is expected by Figs.~\ref{bi-Pmagic} and \ref{bi-P3000}. 
The feature of correlation is least obvious in the combination 
$\varepsilon_{ee} - \varepsilon_{\tau \tau}$ though indicating 
a weak oblique linear dependence. 
For the branch-like structure we will make comments to clarify its 
nature in Sec.~\ref{nonzero-input} because the structure 
is even more prominent with non-zero $\varepsilon$ input. 

The effect of combining the intermediate and the far detectors is remarkable. 
The allowed regions scattered in wide ranges in the top (3000 km) and 
the middle (7000 km) panels combine into a much smaller region in 
the bottom panel in Figs.~\ref{ee-et-tt-piby4}-\ref{ee-em-et-3piby2}. 
We should remark that although the above mentioned 
over-all features remain unchanged for different values of $\delta$, 
the resultant sensitivity to $\varepsilon$'s depends rather strongly on 
the value of CP phase $\delta$ as one can see by comparing between 
Fig.~\ref{ee-et-tt-piby4} and Fig.~\ref{ee-et-tt-3piby2}. 
The problem of $\delta$ dependence of the sensitivity to $\varepsilon$'s 
will be fully addressed in Sec.~\ref{disc-reach}. 

One may ask the question why the scattered regions in the top panel
and the extended region in the middle panel in each column in
Fig.~\ref{ee-et-tt-piby4} and Fig.~\ref{ee-et-tt-3piby2}, and
Fig.~\ref{ee-em-mm-piby4} and Fig.~\ref{ee-em-et-3piby2}, can be
combined to yield such a small region.  The answer is that it is due
to the CP phase $\delta$.  Namely, most of the region of overlap
between the top and the middle panels have mismatch in value of
$\delta$, and hence they do not survive when the two constraints are
combined.  Therefore, keeping the solar $\Delta m^2$ and the KM phase
degree of freedom is the key to the high sensitivity to
$\varepsilon$'s we observe in Figs.~\ref{ee-et-tt-piby4}-\ref{ee-em-et-3piby2}.
The synergy effect that merit us by combining the intermediate 
and the far detectors are even more significant compared with the one in 
identical two detector method for measuring CP violation 
and determining the mass hierarchy \cite{MNplb97,T2KK}. 

Our last comment in this subsection is that a characteristic feature 
that manifest itself in Figs.~\ref{ee-et-tt-piby4}-\ref{ee-em-et-3piby2}
gives us a warning.  Namely, one could significantly overestimate the
sensitivity to detection of non-vanishing NSI by working only with a 
particular single element $\varepsilon_{ee}$ or $\varepsilon_{e
  \tau}$, for example.
By working with two $\varepsilon$'s at the same time one is able 
to recognize the whole structure as given in these figures.\footnote{
  It also raises the question whether the two $\varepsilon$ systems
  are sufficiently generic to reveal the full structure of the systems
  with larger number of $\varepsilon$'s which have a multi-dimensional
  $\chi^2$ manifold.  We are not able to answer the question in this
  paper.  }

\subsection{Sensitivity to $\theta_{13}$ and $\delta$ in the presence of NSI; 
Case of zero input}
\label{13-del-zero}

\begin{figure}[htbp]
\vglue -0.5cm
\begin{center}
\includegraphics[width=1.0\textwidth]{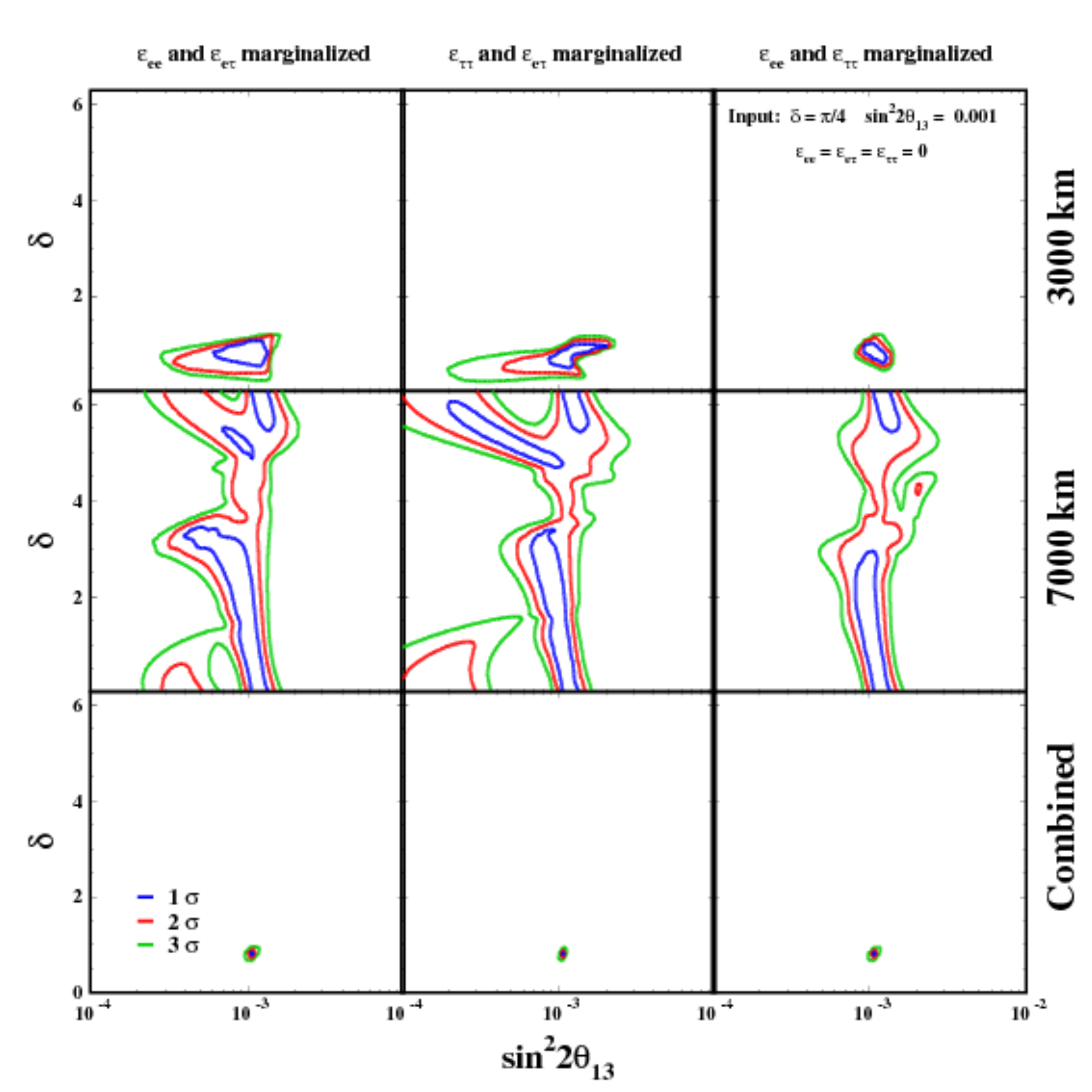}
\end{center}
\vglue -0.5cm
\caption{
Allowed regions projected into the plane of 
$\sin^2 2\theta_{13}$-$\delta$ 
corresponding to the case where 
the input parameters are $\sin^2 2\theta_{13} = 0.001$ 
and $\delta = \pi/4$ and no non-standard interactions
(or all the $\varepsilon$'s are zero),  
for $E_\mu$ = 50 GeV and the baseline of $L=3000$ km (upper panels), 
7000 km (middle horizontal panels) and combination (lower panels). 
The fit was performed by varying freely 4 parameters, 
$\theta_{13}$, $\delta$ and 2 $\varepsilon$'s
where $\varepsilon_{ee}$ and $\varepsilon_{e\tau}$ 
are marginalized (left panels), 
$\varepsilon_{\tau\tau}$ and $\varepsilon_{e\tau}$ 
are marginalized (middle panels) 
and $\varepsilon_{ee}$ and $\varepsilon_{\tau\tau}$ 
are marginalized (right panels).  
The same input and fitting parameters as in 
Fig.~\ref{ee-et-tt-piby4} but projections 
are made into the different parameter space. 
}
\label{nsi-th13-del-piby4-tau} 
\end{figure}

\begin{figure}[htbp]
\vglue -0.5cm
\begin{center}
\includegraphics[width=1.0\textwidth]{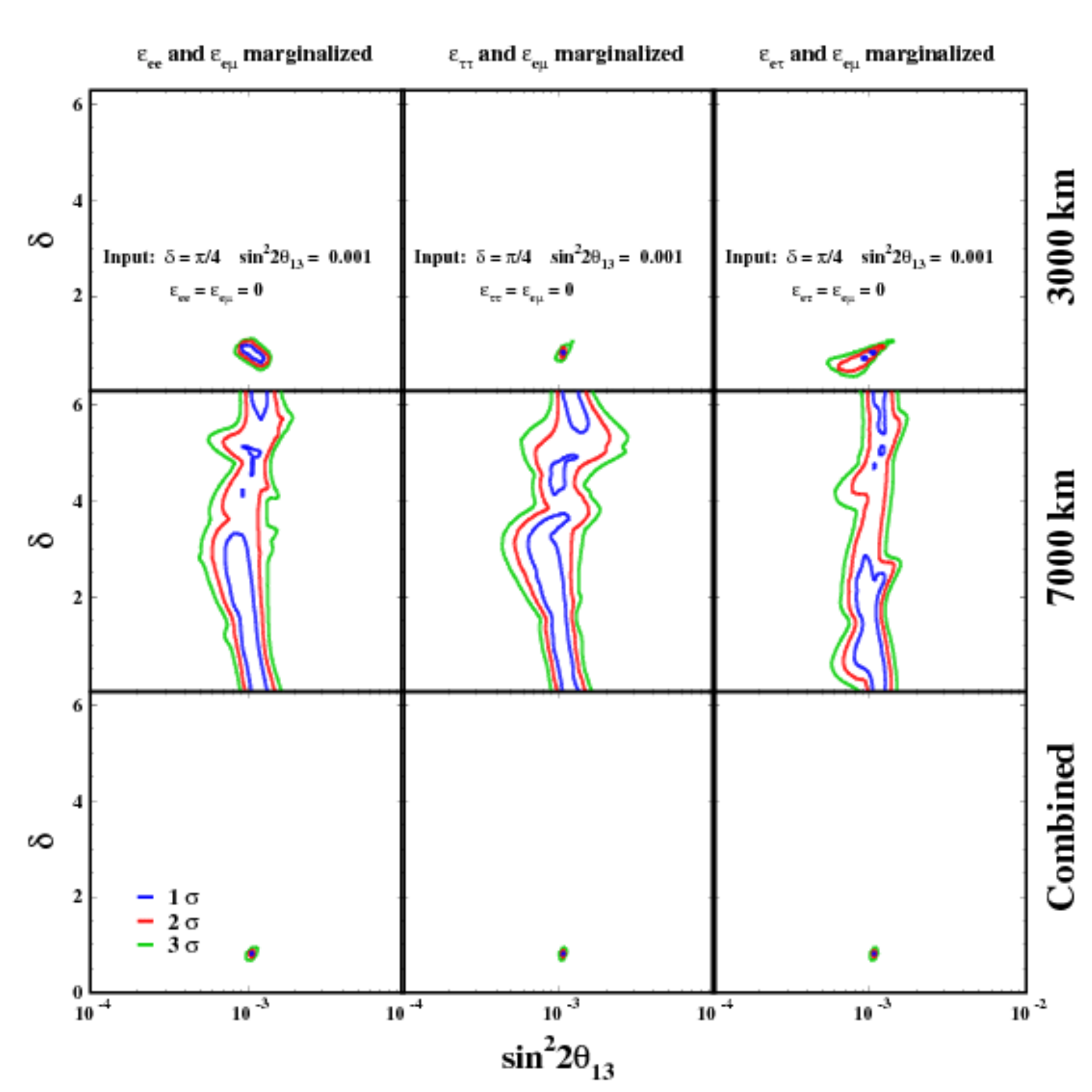}
\end{center}
\vglue -0.5cm
\caption{The same as in Fig.~\ref{nsi-th13-del-piby4-tau}  but 
for different combination of 2 $\varepsilon$'s 
to which the fit to $\sin^2 2\theta_{13}$ and $\delta$ is marginalized;  
$\varepsilon_{ee}$-$\varepsilon_{e\mu}$ 
(left panels), $\varepsilon_{\tau\tau}$-$\varepsilon_{e\mu}$ (middle panels)
and $\varepsilon_{e\mu}$-$\varepsilon_{e\tau}$ (right panels). 
}
\label{nsi-th13-del-piby4-mu}
\end{figure}

\begin{figure}[htbp]
\vglue -0.5cm
\begin{center}
\includegraphics[width=1.0\textwidth]{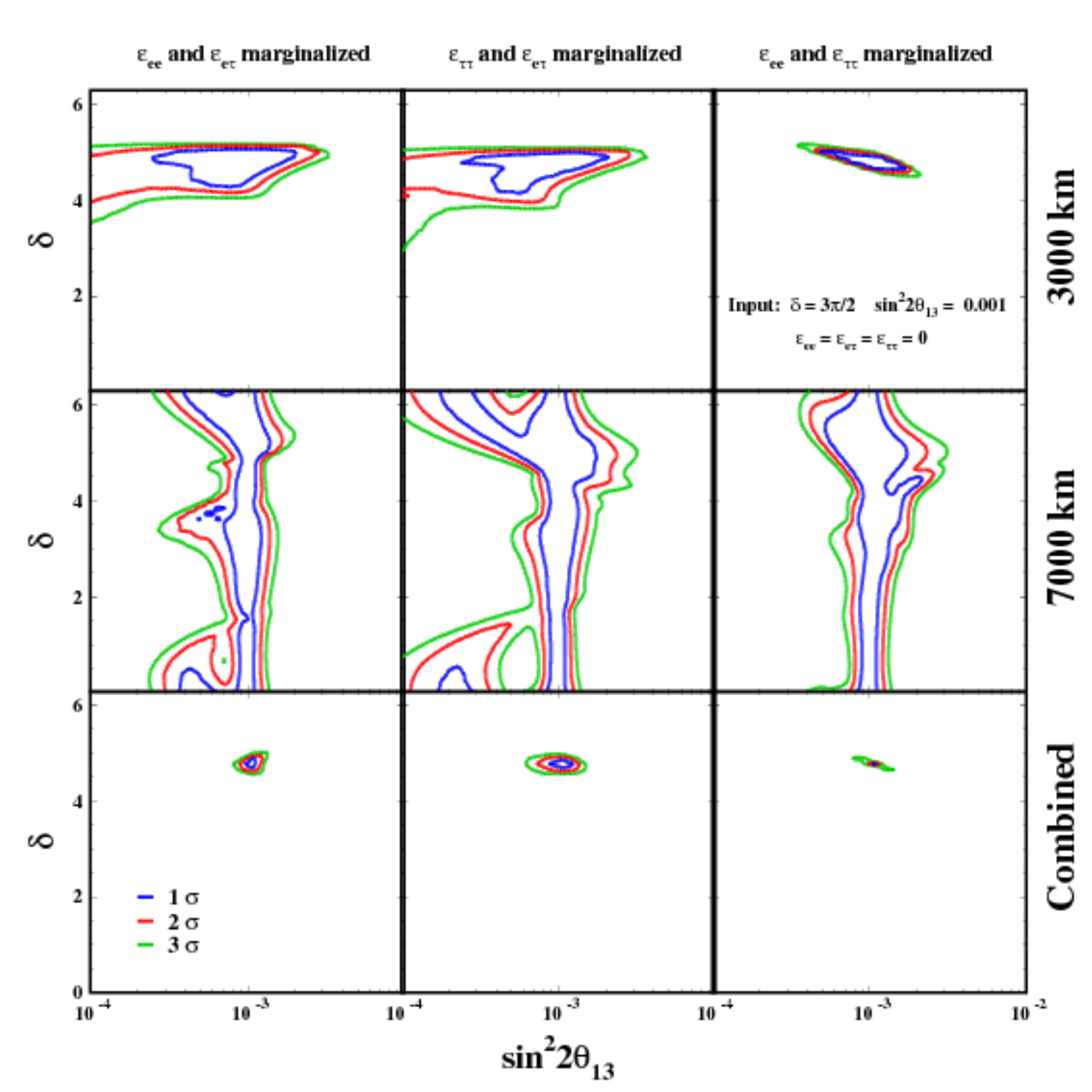}
\end{center}
\caption{The same as in Fig.~\ref{nsi-th13-del-piby4-tau}  but with 
$\delta = 3\pi/2$. 
}
\label{nsi-th13-del-3piby2-tau}
\end{figure}

\begin{figure}[htbp]
\vglue -0.5cm
\begin{center}
\includegraphics[width=1.0\textwidth]{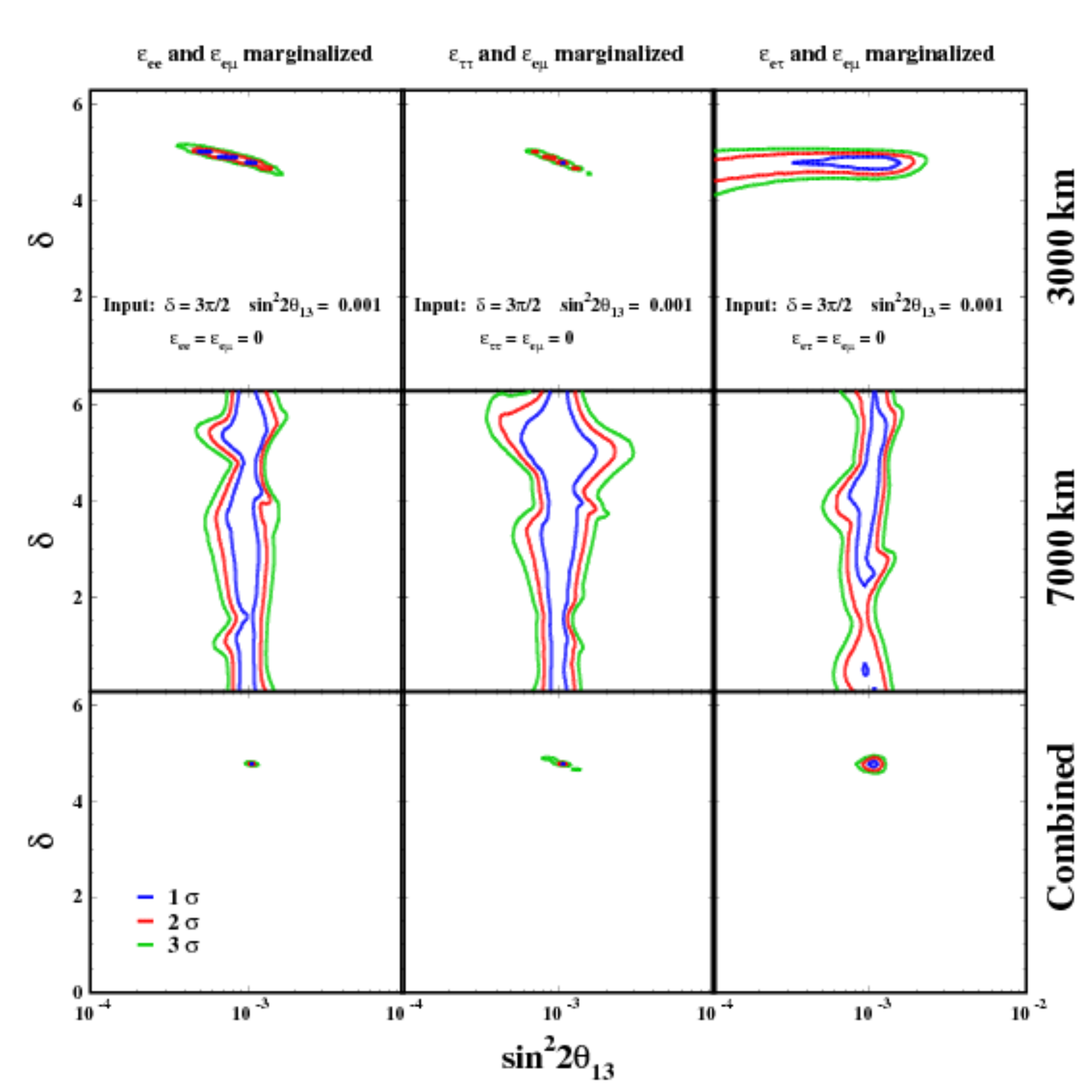}
\end{center}
\caption{The same as in Fig.~\ref{nsi-th13-del-piby4-mu}  but with 
$\delta = 3\pi/2$. 
}
\label{nsi-th13-del-3piby2-mu}
\end{figure}

We now examine the sensitivities to $\theta_{13}$ and $\delta$ in the
presence of NSI.  In Fig.~\ref{nsi-th13-del-piby4-tau} through
Fig.~\ref{nsi-th13-del-3piby2-mu} we present the sensitivities to
$\theta_{13}$ and $\delta$ by taking the same input value of
$\theta_{13}$, $\sin^2 2\theta_{13} = 10^{-3}$.
Figs.~\ref{nsi-th13-del-piby4-tau} and \ref{nsi-th13-del-piby4-mu} are
for $\delta=\pi/4$, whereas Figs.~\ref{nsi-th13-del-3piby2-tau} and
\ref{nsi-th13-del-3piby2-mu} are $\delta=3\pi/2$.  The organization of
the figures is the same as in Figs.~\ref{ee-et-tt-piby4}-\ref{ee-em-et-3piby2}.

Let us look at the top panel for the detector at $L=3000$ km.  We clearly
observe the phenomenon of ``confusion'' in the presence of NSI, in
particular for the case of $\delta=3\pi/2$.  Namely, NSI can mimic the
effect of non-zero $\theta_{13}$ so that the allowed region extends to
a very small value of $\theta_{13}$.  However, this feature is
strongly perturbed by the measurement at $L=7000$ km.  Because it is
highly sensitive to the effects of NSI, it helps to resolve the
confusion between NSI and $\theta_{13}$.  After combining informations
of the intermediate and the far detectors a tiny region in $\sin^2
2\theta_{13} - \delta$ space results, as one can observe in the bottom
panels in Fig.~\ref{nsi-th13-del-piby4-tau}  through
Fig.~\ref{nsi-th13-del-3piby2-mu}.
Thus, the problem of NSI - $\theta_{13}$ confusion can be solved 
by the two detector setting.\footnote{
The ``confusion theorem'' stated in \cite{confusion2} refers to the situation 
in which we have both NSI at source and in propagation and they are 
related with each other in a specific way. Therefore, 
our discussion in this paper does not affect the validity of this theorem.
}

\subsection{How do the sensitivities depend on $\theta_{13}$?}
\label{0.0001}

\begin{figure}[htbp]
\vglue -0.5cm
\begin{center}
\includegraphics[width=1.0\textwidth]{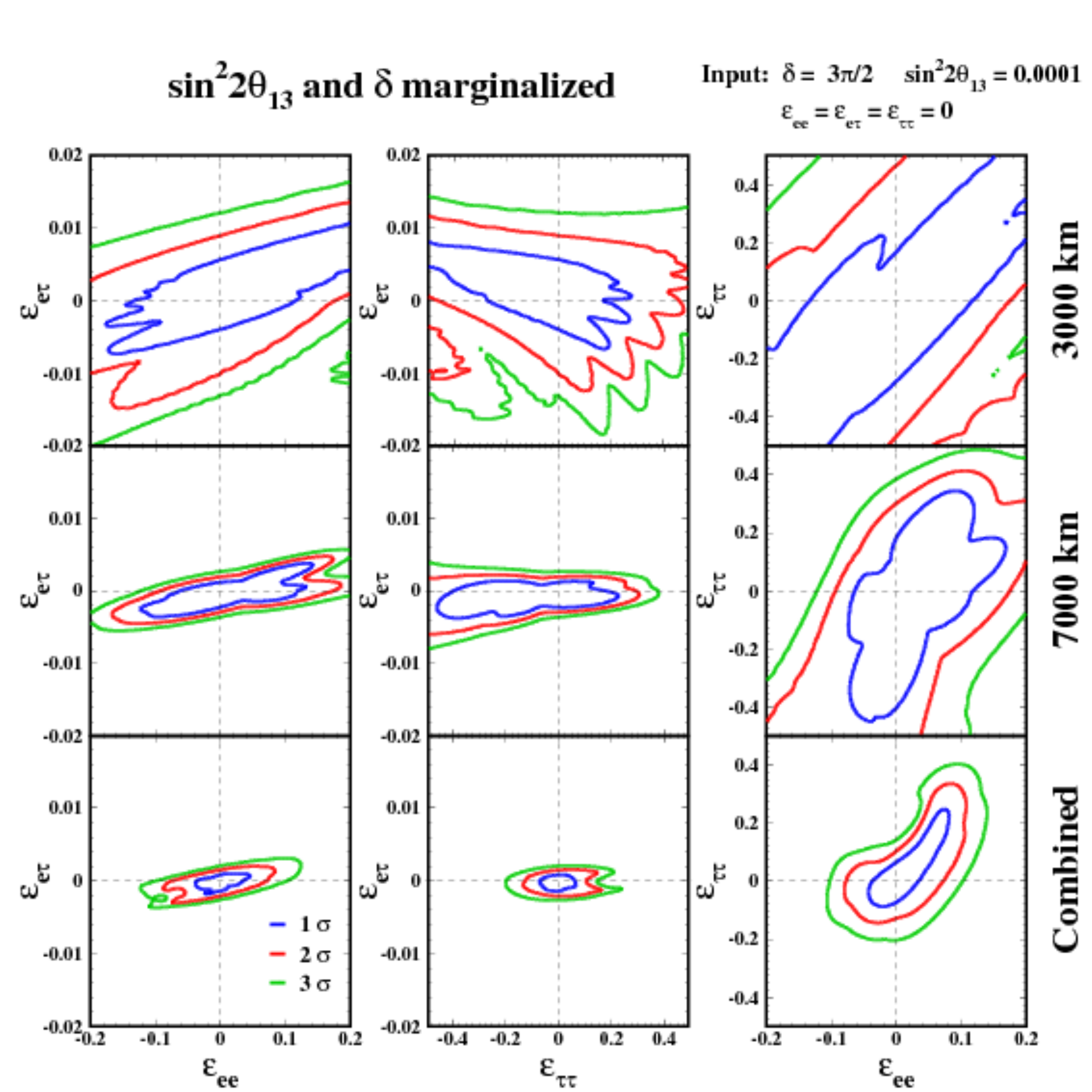}
\end{center}
\caption{The same as in Fig.~\ref{ee-et-tt-3piby2} but with 
$\sin^2 2\theta_{13}=0.0001$.  
}
\label{ee-et-tt-3piby2-0.0001}
\end{figure}

\begin{figure}[htbp] 
\vglue -0.5cm
\begin{center}
\includegraphics[width=1.0\textwidth]{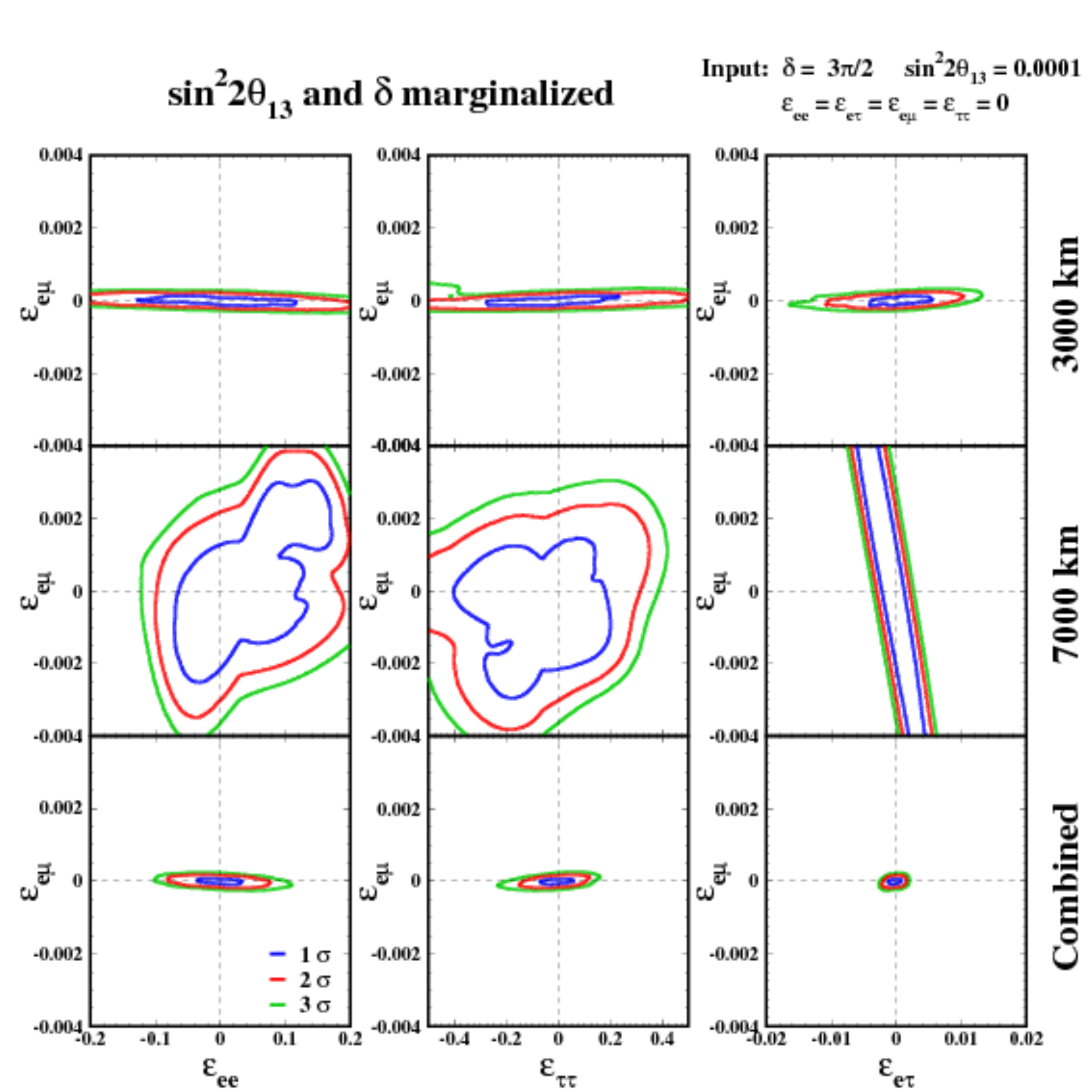}
\end{center}
\caption{The same as in Fig.~\ref{ee-em-et-3piby2} 
but with $\sin^2 2\theta_{13} = 0.0001$.
}
\label{ee-em-mm-3piby2-0.0001}
\end{figure}

\begin{figure}[htbp]
\vglue -0.5cm
\begin{center}
\includegraphics[width=1.0\textwidth]{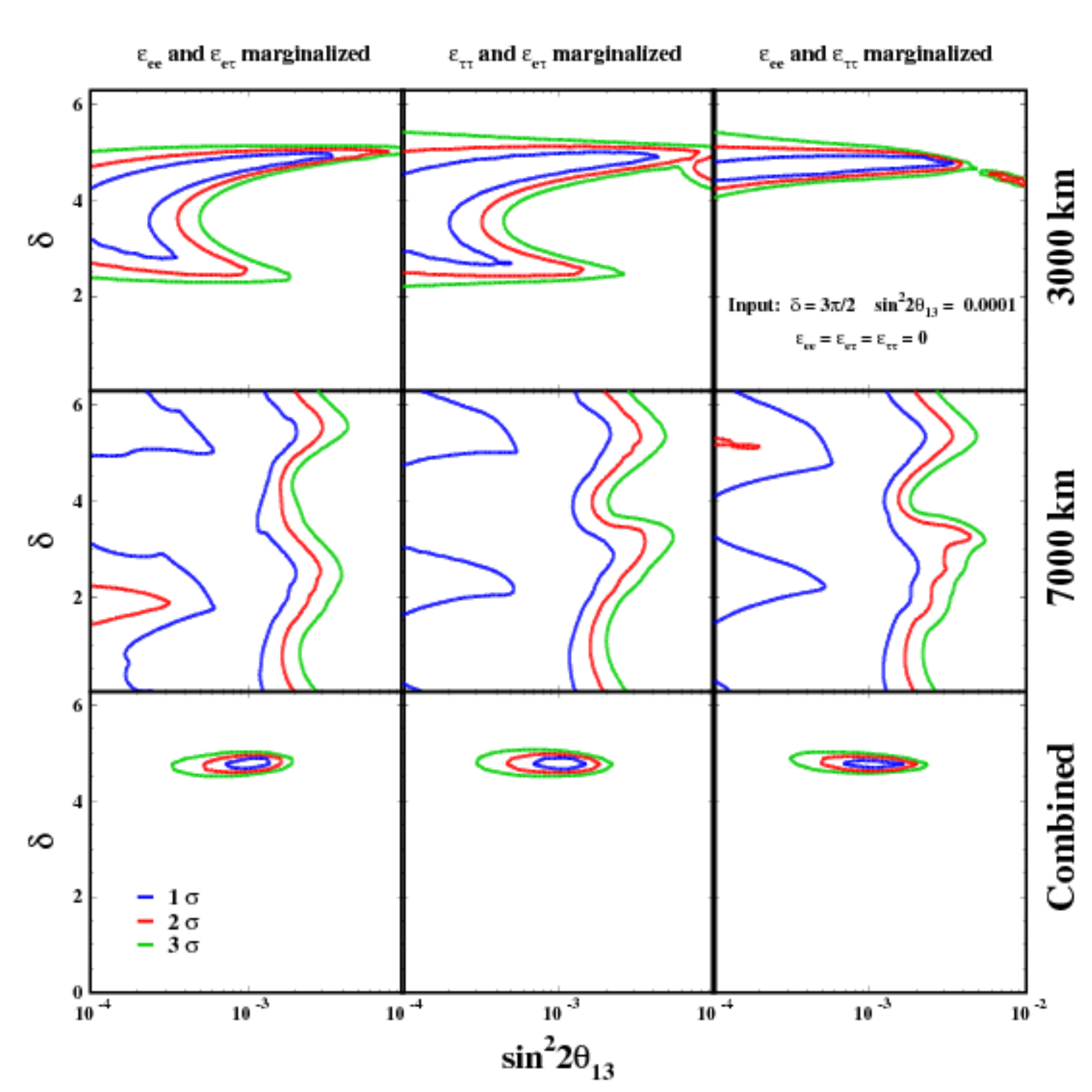}
\end{center}
\caption{The same as in Fig.~\ref{nsi-th13-del-3piby2-tau}  but with 
$\sin^2 2\theta_{13}=0.0001$.
}
\label{nsi-th13-del-3piby2-tau-0.0001}
\end{figure}

\begin{figure}[htbp]
\vglue -0.5cm
\begin{center}
\includegraphics[width=1.0\textwidth]{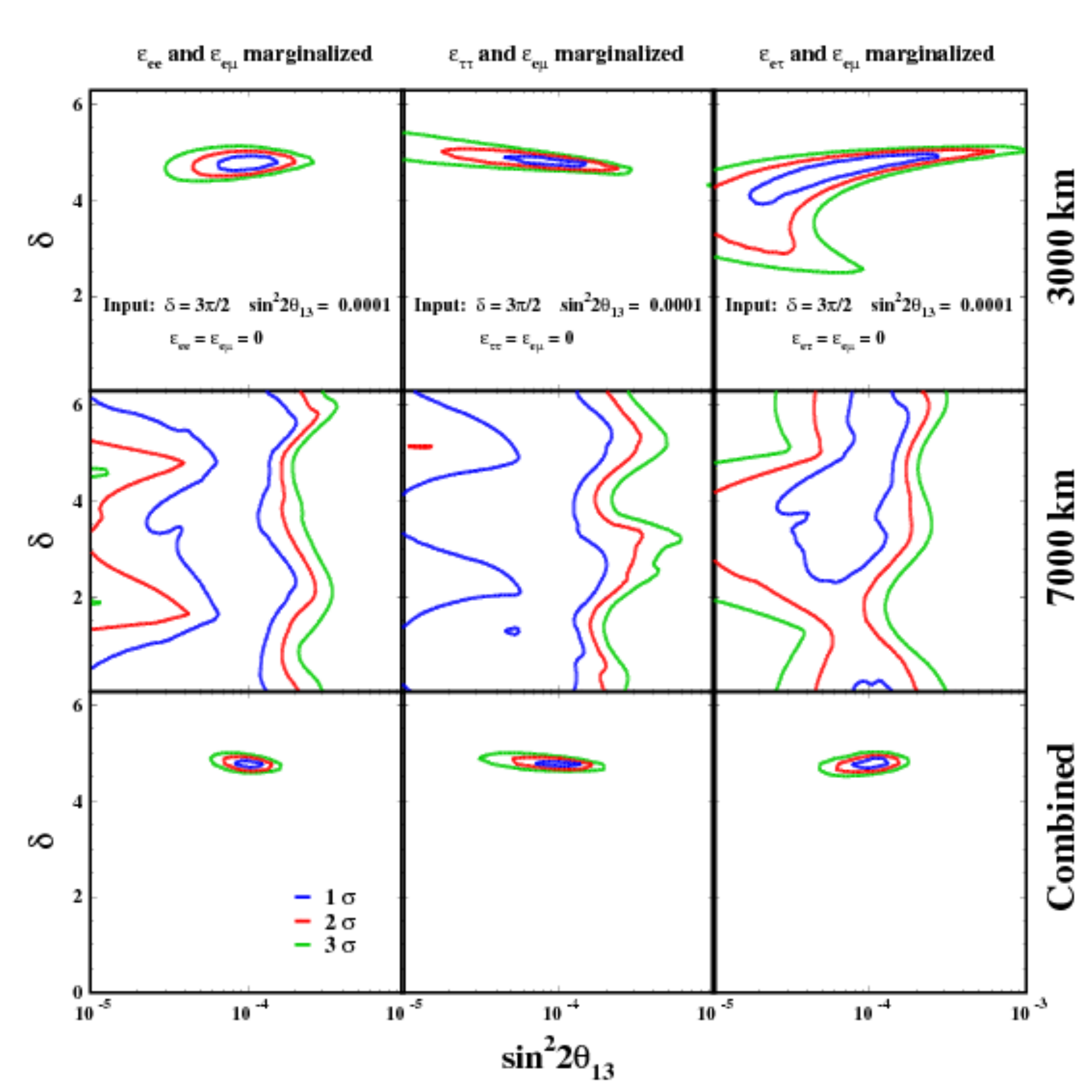}
\end{center}
\caption{The same as in Fig.~\ref{nsi-th13-del-3piby2-mu}  but with 
$\sin^2 2\theta_{13}=0.0001$. 
}
\label{nsi-th13-del-3piby2-mu-0.0001}
\end{figure}

We discuss briefly how the sensitivities change when we take 
a smaller value of $\theta_{13}$, for example, 
$\sin^2 2 \theta_{13}=0.0001$. 
We present our results only for $\delta = 3\pi/2$ for reasons 
of limitation of space but make some comments on the other cases.  
In Figs.~\ref{ee-et-tt-3piby2-0.0001} and
\ref{ee-em-mm-3piby2-0.0001} we show the sensitivities to NSI
parameters which may be compared to the corresponding figures,
Figs.~\ref{ee-et-tt-3piby2} and \ref{ee-em-et-3piby2}, for the case
$\sin^2 2 \theta_{13}=0.001$.
To our surprise, the over-all features of the allowed regions are similar 
between the two cases. The only notable changes are that: 
(1) in the $\varepsilon_{e e} - \varepsilon_{\tau \tau}$ system 
the sensitivities to these $\varepsilon$'s becomes worse by a 
factor of 3-4, but 
(2) in all the three systems which involve $\varepsilon_{e \mu}$,  
$\varepsilon_{e e} - \varepsilon_{e \mu}$, 
$\varepsilon_{\tau \tau} - \varepsilon_{e \mu}$, and 
$\varepsilon_{e \tau} - \varepsilon_{e \mu}$, 
the sensitivities become {\em better} for $\sin^2 2 \theta_{13}=0.0001$. 
The similar worsening or improvement of the sensitivities to NSI 
are observed in other values of $\delta$. 
They become worse in some cases, in particular, in the 
$\varepsilon_{e e} - \varepsilon_{\tau \tau}$ system 
with $\delta = \pi/4$ and $\pi/2$ at $L=3000$ km. 
The sensitivities to $\varepsilon_{e e}$ and $\varepsilon_{\tau \tau}$
improve by almost a factor of 2 at $\delta=\pi$ with $\sin^2 2
\theta_{13}=0.0001$ compared to the case of 0.001.  The sensitivity to
$\varepsilon_{e \tau}$ gets improved by about a factor of 2 in the
$\varepsilon_{e \mu} - \varepsilon_{e \tau}$ system at $\delta=\pi/4$.
Thus, most of the improvement occur in the systems which involve
$\varepsilon_{e \mu}$.  The reason is that the sensitivity to
$\varepsilon_{e \mu}$ is particularly good at $L=3000$ km and it makes
the synergy effect even stronger.

We notice that the branch structure we saw in Sec.~\ref{NSI-zero} is
also rounded off, and the island structure becomes less prominent at
$\sin^2 2 \theta_{13}=0.0001$.
The smoothing of the contour and the improvement of the sensitivity 
may be due to the fact that at such very small $\theta_{13}$ 
the effect of NSI becomes important compared to the 
standard oscillation effect.

On the other hand, the sensitivity to $\theta_{13}$ and $\delta$
definitely becomes worse when we go down to $\sin^2 2
\theta_{13}=0.0001$ as shown in Figs.~\ref{nsi-th13-del-3piby2-tau-0.0001} and
\ref{nsi-th13-del-3piby2-mu-0.0001}.  First of all, the problem of
$\theta_{13}$-NSI confusion becomes severer as one can see by
comparing Figs.~\ref{nsi-th13-del-3piby2-tau-0.0001} and
\ref{nsi-th13-del-3piby2-mu-0.0001} to
Figs.~\ref{nsi-th13-del-3piby2-tau} and \ref{nsi-th13-del-3piby2-mu}.
A new allowed region emerges in the 
$\varepsilon_{e e} - \varepsilon_{e \tau}$ and 
$\varepsilon_{\tau \tau} - \varepsilon_{e \tau}$ systems at $L=3000$ km. 
Due to lack of statistics at the small $\theta_{13}$ 
sensitivity to $\theta_{13}$ is significantly reduced at $L=7000$ km, 
leading to the visible loss of the sensitivities.

In summary, neutrino factory with two detector setup at $L=3000$ and
7000 km is more resistant to small values of $\theta_{13}$ as a
discovery machine for NSI, rather than as a machine for precision
measurement of $\theta_{13}$ and $\delta$.

\subsection{Comparing the sensitivities to $\theta_{13}$ and $\delta$ for 
cases with and without NSI}

\begin{figure}[htbp]
\begin{center}
\includegraphics[width=0.9\textwidth]{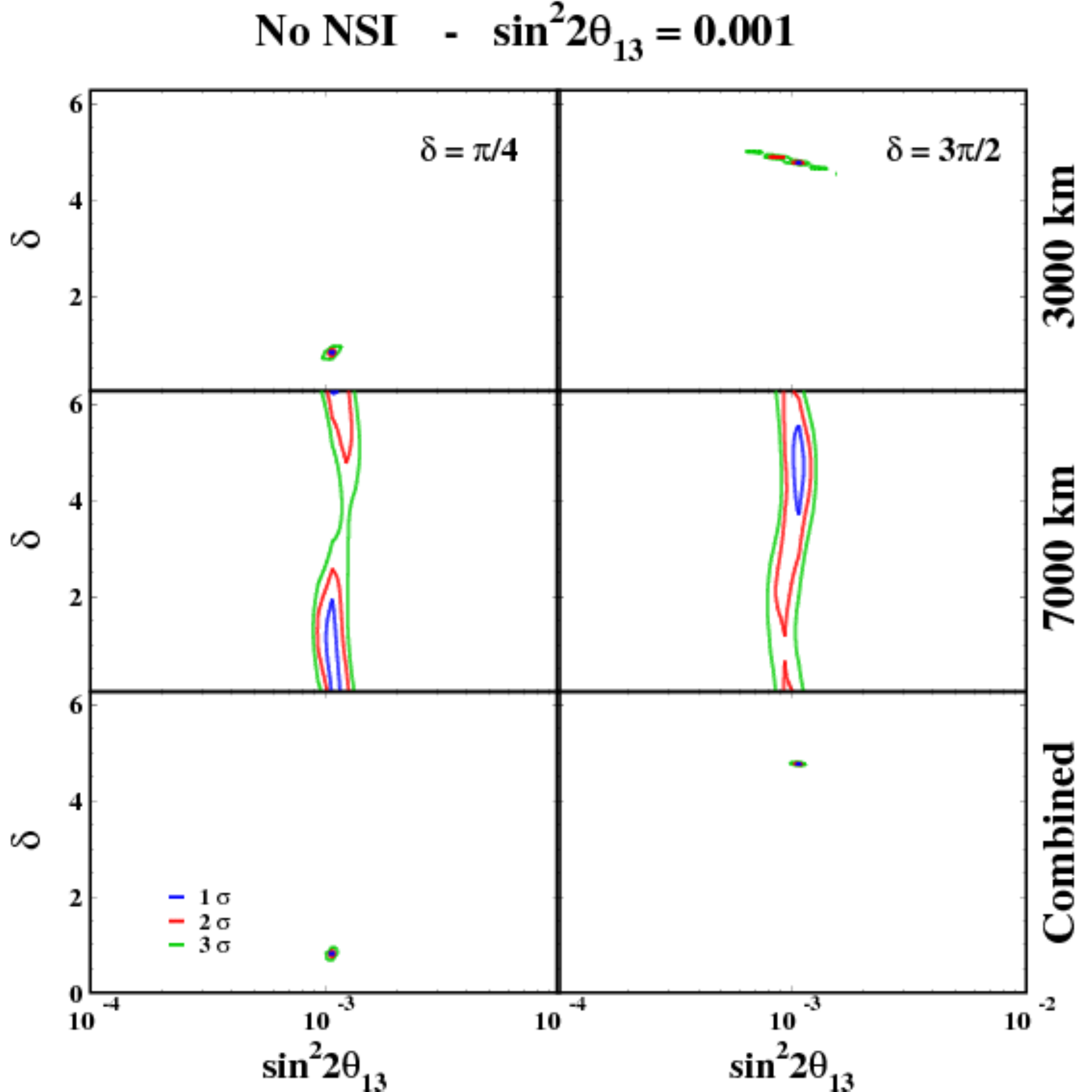}
\end{center}
\caption{Allowed regions in the plane $\sin^2 2\theta_{13}-\delta$ for
  the standard oscillation case (no NSI) for the input values $\sin^2
  2\theta_{13}=0.001$, $\delta=\pi/4$ (left panels) and
  $\delta=3\pi/2$ (right panels).  The allowed regions were computed
  for 2 DOF.  The experimental conditions are the same as for 
  Fig.~\ref{nsi-th13-del-piby4-tau}.  }
\label{standard}
\end{figure}

It may be worthwhile to compare the sensitivities to $\theta_{13}$ and
$\delta$ for cases without NSI to the one with NSI using the same
machinery used in our foregoing analysis.  In Fig.~\ref{standard} we
present the results without NSI.  The results are to be compared with
those in Figs.~\ref{nsi-th13-del-piby4-tau}-\ref{nsi-th13-del-3piby2-mu}.
The resultant sensitivities to $\theta_{13}$ and $\delta$ are
extremely good compared to those obtained for the cases with NSI.  
We will quantify this comparison in a more extensive way in
Sec.~\ref{disc-reach} where we will cover the whole space of interest
in $\theta_{13}$ and $\delta$.

\subsection{Effect of background and systematic errors} 
\label{systematic}

To have a rough idea of the possible effects of the background 
and the systematic uncertainties to the sensitivities 
we have repeated the computation with these effects at 
$\delta=\pi/4$ for various values of $\theta_{13}$ in 
$\varepsilon_{ee}$-$\varepsilon_{e\tau}$, 
$\varepsilon_{\tau\tau}$-$\varepsilon_{e\tau}$, and 
$\varepsilon_{ee}$-$\varepsilon_{\tau\tau}$ systems, 
which correspond to Fig.~\ref{ee-et-tt-piby4}, 
Fig.~\ref{nsi-th13-del-piby4-tau}, and $\delta=\pi/4$ counterpart of 
Fig.~\ref{ee-et-tt-3piby2-0.0001} (not shown). 
We estimate the number of background events by using the signal to 
noise ratio calculated in Ref.~\cite{dydak}, and rescaling the background 
to the number of useful muon decays used in this paper. 
We assign 20\%  uncertainty to the number of background events. 
We also take 2.5\% as the total systematic uncertainty on the 
measurement. 

For $\sin^2 2\theta_{13}=0.001$, we find that the introduction of 
systematic uncertainties and background lead to decrease in the
sensitivity to $\varepsilon_{e\tau}$ by $\simeq$ 30\%.  The worsening
of the sensitivity to $\varepsilon_{ee}$ and $\varepsilon_{\tau\tau}$
depends upon the sub-system being within the ranges 30\%-60\% and
40\%-50\%, respectively. 
The effect of systematic errors and background is less prominent 
at $\sin^2 2\theta_{13}=0.005$ but becomes larger at 
$\sin^2 2\theta_{13}=0.0001$, making the sensitivity a factor of 2 worse. 

On the other hand, the sensitivity to $\delta$  and 
$\sin^2 2\theta_{13}$ would be worsen by 30\%-40\% and 50\%, 
respectivelly, at  $\sin^2 2\theta_{13}=0.001$, more or less  
independently of  the sub-system treated and on the $\varepsilon$'s 
discussed.  An interesting feature of the sensitivity to $\delta$ 
and $\theta_{13}$ is that the effect of systematic errors 
diminishes as $\theta_{13}$ becomes small, and is at $\sim 20$\% (30\%) 
level for $\delta$ and $\sin^2 2\theta_{13}$, respectivelly,  
 at $\sin^2 2\theta_{13}=0.0001$.  Therefore, the impact of including 
the systematic to the sensitivity of $\delta$ and $\theta_{13}$ is 
limited in size, which {\em is} good news.  All numbers were 
evaluated at 3 $\sigma$ CL.

Though we continue to ignore the systematic uncertainties and the 
background in the rest of the analysis in this paper, the reader should  
remind that the estimated sensitivities to NSI, $\delta$, and 
$\theta_{13}$ have uncertainties at the level quoted above. 
To completely remove the uncertainties requires precise knowledges 
of the performance of the detector. 

\section{Accuracies of Determination of NSI, $\theta_{13}$, and $\delta$} 
\label{nonzero-input}

We now discuss the question of how well the magnitude of non-vanishing
NSI can be determined by the two detector setting, and at the same
time to what extent the measurement of $\theta_{13}$ and $\delta$ can be
affected by non-zero input values of NSI.  
In this section, we examine the particular systems: 
$\varepsilon_{e e} - \varepsilon_{e \tau}$, 
$\varepsilon_{\tau \tau} - \varepsilon_{e \tau}$, and 
$\varepsilon_{e e} - \varepsilon_{\tau \tau}$.
We do not discuss the case which involve $\varepsilon_{e \mu}$. 
The reason is that we want to take the NSI input values well below 
the current bounds  presented Eq.~(\ref{bound}).
The input value of $\varepsilon_{e \mu}$ which is comparable to 
the present constraint will not affect the sensitivity. 
In Sec.~\ref{disc-reach}, we will present more global informations. 
We work with the particular input values of $\theta_{13}$, $\delta$, 
and the $\varepsilon$ parameters; 
$\sin^2 2 \theta_{13} = 10^{-3}$, 
$\delta = 3\pi/2$, 
$\varepsilon_{e \tau} = 0.01$, 
$\varepsilon_{e e} = 0.1$, and 
$\varepsilon_{\tau \tau} = 0.2$. 
But, the features are not so different for other input values.

\subsection{Sensitivity to NSI, $\theta_{13}$, and $\delta$; Case of non-zero 
input of NSI}

Roughly speaking, the sensitivities to NSI, $\theta_{13}$, and $\delta$ 
are not affected so much by the non-zero input values of $\varepsilon$'s. 
This is because, even in the case of vanishing input, we freely vary 
them in fitting the data, and the presence of these extra degrees of 
freedom of varying over the NSI parameters is of key importance to 
determine (or affect) the sensitivities. 
Since the results are similar to the previous case, where NSI input were set 
to  zero, we only present the figures for the case $\delta=3\pi/2$. 
Generally speaking, the sensitivity is worse than for the case 
$\delta=\pi/4$. 
More or less one can guess what would be the general characteristics of 
this case from the figures with zero input,  
Figs.~\ref{ee-et-tt-piby4}, \ref{ee-em-mm-piby4}, 
\ref{nsi-th13-del-piby4-tau}  and \ref{nsi-th13-del-piby4-mu}. 

\begin{figure}[htbp]
\begin{center}
\includegraphics[width=1.0\textwidth]{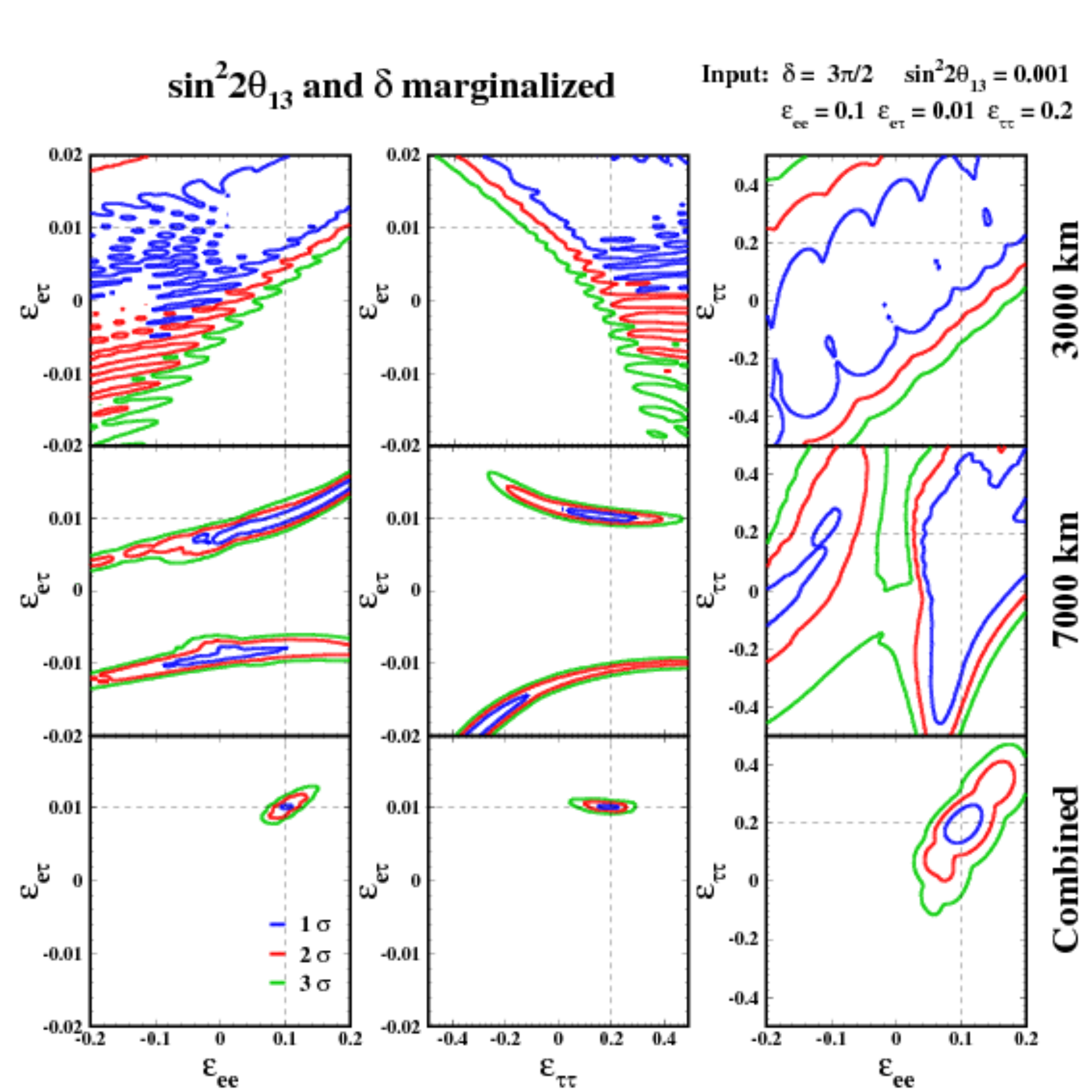}
\end{center}
\caption{
The same as in Fig.~\ref{ee-et-tt-3piby2} 
but for non-vanishing input values of $\varepsilon$; 
$\varepsilon_{ee} = 0.1$,  $\varepsilon_{e\tau} = 0.01$ and  
$\varepsilon_{\tau\tau} = 0.2$. 
We note that only the input values of 2 $\varepsilon$'s are 
set to be non-zero at the same time.  
The thin dashed lines indicate the corresponding non-zero values of 
$\varepsilon_{\alpha \beta}$ for each panel.
}
\label{ee-et-tt-3piby2-nonzero}
\end{figure}

We first discuss the accuracy of the determination of NSI parameters. 
Presented in Fig.~\ref{ee-et-tt-3piby2-nonzero} is the sensitivities to 
NSI parameters with the non-zero inputs. 
We notice that the features of the allowed regions are essentially 
the same as in the zero input case shown in 
Fig.~\ref{ee-et-tt-3piby2}. 
(About the two split islands in the middle panels see below.) 
We note that the accuracy of the determination of the $\varepsilon$
parameters varies case by case.  Most significantly, the size of
allowed region of $\varepsilon_{e \tau}$, $\varepsilon_{e e}$, and
$\varepsilon_{\tau \tau}$ become worse by more than a factor 2 in
$\varepsilon_{e e} - \varepsilon_{e \tau}$ and $\varepsilon_{\tau \tau} -
\varepsilon_{e \tau}$ systems.
However, the accuracy of the determination of $\varepsilon_{e \tau}$ and 
$\varepsilon_{\tau \tau}$ at $\delta= \pi/4$ (not shown) improves 
by a factor of $\simeq$ 2 with the non-zero input values of NSI.

\begin{figure}[htbp]
\begin{center}
\includegraphics[width=1.0\textwidth]{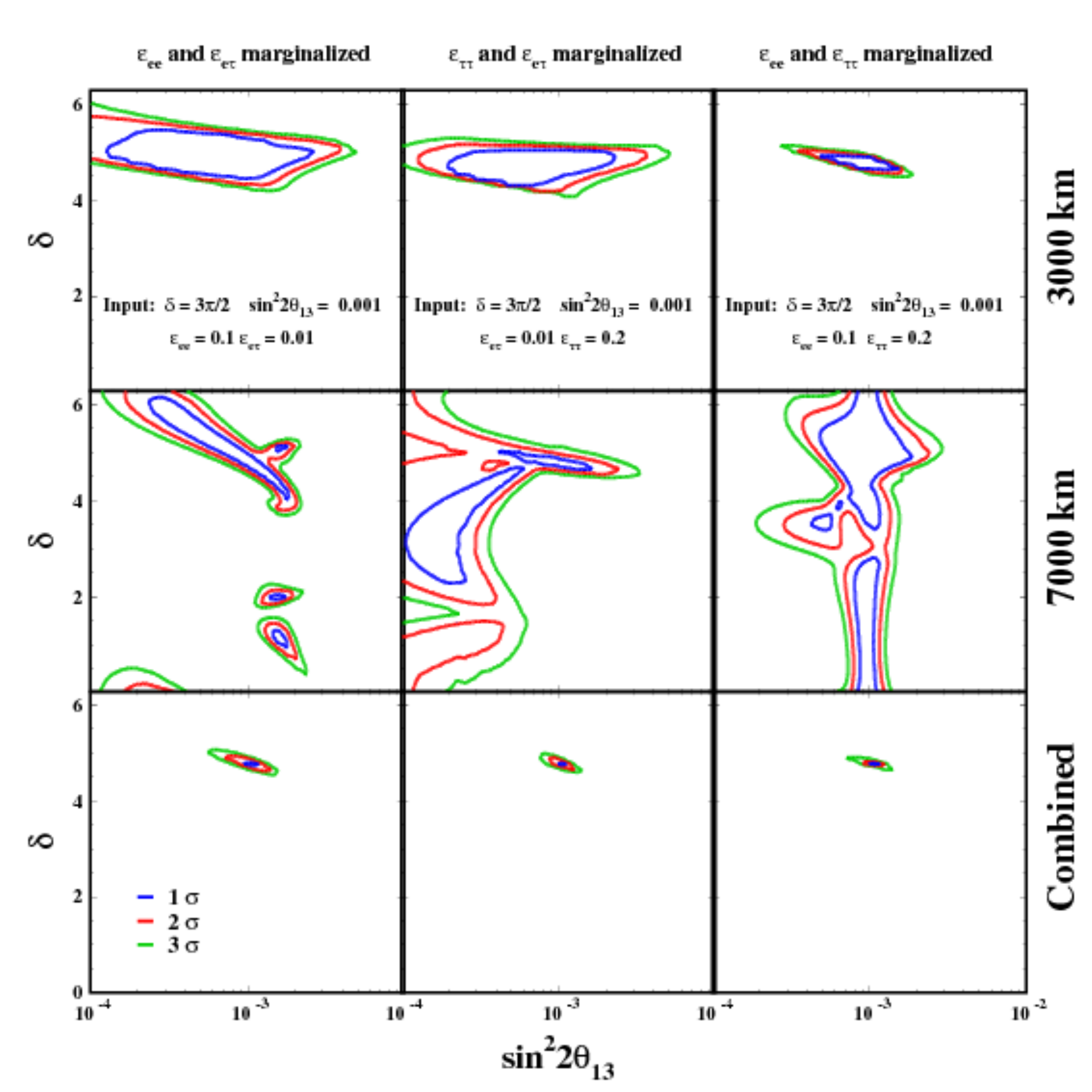}
\end{center}
\caption{
The same as in Fig.~\ref{nsi-th13-del-3piby2-tau} 
but for non-vanishing input values of $\varepsilon$; 
$\varepsilon_{ee} = 0.1$,  $\varepsilon_{e\tau} = 0.01$ and 
$\varepsilon_{\tau\tau} = 0.2$. 
}
\label{nsi-th13-del-3piby2-tau-nonzero}
\end{figure}

Next, we discuss the accuracy of the measurement of $\theta_{13}$
and $\delta$ in the presence of non-vanishing input values of NSI.
Fig.~\ref{nsi-th13-del-3piby2-tau-nonzero} serves for this purpose.
Again the features of Fig.~\ref{nsi-th13-del-3piby2-tau-nonzero} are
similar to those of Fig.~\ref{nsi-th13-del-3piby2-tau}.
We, however, notice some new characteristics.
The $\theta_{13}$-NSI confusion at $L=3000$ km is severer for 
$\delta=3\pi/2$, but it is milder for $\delta=\pi/4$ (not shown) 
in the present case compared to zero $\varepsilon$ input. 
Therefore, the degree of confusion depends very much on $\delta$. 
Also the resultant accuracy of the determination of $\theta_{13}$ and $\delta$, 
after the two detectors are combined, depends upon which $\varepsilon$ is 
tuned on and also on $\delta$ though only mildly. 
For $\delta=3\pi/2$, the uncertainty on $\sin^2 2 \theta_{13}$ is
smaller (larger) for the case of non-zero input value of the NSI parameters 
than for the zero-input case in the $\varepsilon_{\tau \tau} - \varepsilon_{e \tau}$
($\varepsilon_{ee} - \varepsilon_{e \tau}$) system.
It is interesting (and encouraging) to observe that non-zero input
values of NSI essentially do not disturb the sensitivities to the
determination of the NSI parameters but also the sensitivities to  
$\theta_{13}$ and $\delta$.

\subsection{Parameter degeneracies with and without NSI} 
\label{degeneracy}

In this subsection we want to make some remarks on the degeneracies
associated with the measurement of NSI parameters as well as $\theta_{13}
$ and $\delta$.
In Fig.~\ref{ee-et-tt-3piby2-nonzero}, we observe two clearly
separated islands in the regions allowed by the far detector
measurement (middle panels) for the $\varepsilon_{\tau \tau} -
\varepsilon_{e \tau}$ and the $\varepsilon_{ee} - \varepsilon_{e
  \tau}$ systems.  A similar structure exists also in the combinations 
involving $\varepsilon_{e \mu}$.
There exits a degeneracy of solutions of $\varepsilon$'s.  It is easy
to understand the cause of the degeneracy; It is due to the invariance
under sign change of $\varepsilon_{e \tau}$ that can be absorbed by
rotation of $\delta$ by $\pi$.  The fact that the symmetry exists only
at the magic baseline can be easily seen by comparing
Eqs.~(\ref{Penu-2nd_etau}) and (\ref{Pemu_magic-etau}).  This is also
true in the corresponding formula with $\varepsilon_{e \mu}$ that one
can derive from Eq.~(\ref{Penu-2nd_emu}).  Since $\delta$ is
marginalized the two clones appear in the allowed regions in the
$\varepsilon_{\tau \tau} - \varepsilon_{e \tau}$ and 
$\varepsilon_{ee} - \varepsilon_{e \tau}$ space.\footnote{
  If one wants to interpret this symmetry as the one which exist in
  the truncated system in which $\Delta m^2_{21}$ is artificially
  switched off, it is a discrete version of the two phase degeneracy
  discussed in Sec.~\ref{2epsilon}.  }

When the measurement at the two detectors, intermediate and far, are
combined we obtain a unique allowed region; The degeneracy is
resolved.  It must be the case because there is no such symmetry in
the whole system, see Eq.~(\ref{Penu-2nd_etau}) or (\ref{Penu-2nd_emu}).  
But, it is nice to see that it actually occurs in the concrete two detector
setting adopted in this paper.

We want to make brief remarks on the conventional parameter 
degeneracy \cite{intrinsic,MNjhep01,octant} 
which would affect the sensitivity to the lepton mixing parameter. 
It might also affect the sensitivity to NSI 
through its effect to $\theta_{13} $ and $\delta$. 
First of all, there is no $\theta_{23}$ octant degeneracy because 
$\theta_{23}$ is held fixed to $\pi/4$ in our analysis. 
With regard to the four-fold $\theta_{13} - \delta$ degeneracy 
duplicated by the unknown sign of $\Delta m^2_{31}$ we have no 
hint for its existence in our analysis, apart from some limited cases at 
$\sin^2 2\theta_{13}=10^{-4}$ which might be affected by it. 
The reason for this observation is that the clone solutions have a 
significantly different value of $\delta$ from the true solution 
(very roughly speaking $\delta_{\rm clone} \approx \pi - \delta_{\rm true}$), 
in both the intrinsic and the sign-$\Delta m^2_{31}$ degeneracies. 
Nonetheless, from the figures we presented (and also in all those not shown) 
these is no such clone region. 
The reasons are: 
(1) The intrinsic degeneracy is resolved by the spectrum information\footnote{
We have confirmed that the intrinsic degeneracy survives the rate only analysis.  
}, and (2) With long enough baselines the mass hierarchy is
determined.
Therefore, we suspect that the conventional parameter degeneracy plays
minor role, if any, in our sensitivity analysis, except for the one
associated with $\theta_{23}$ which we do not consider in this work.

\section{Discovery reach to NSI, $\theta_{13}$ and $\delta$} 
\label{disc-reach}

In this section we try to summarize the sensitivities to 
NSI, $\theta_{13}$ and $\delta$ that can be achieved by 
a neutrino factory with the intermediate-far detector setting. 
However, in looking for ways to present the sensitivities, 
we recognized that the sensitivities depend very much 
on the input values of $\delta$ and $\theta_{13}$. 
Therefore, we need to show the sensitivities as a function of both 
$\delta$ and $\theta_{13}$ simultaneously. 

\begin{figure}[htbp]
\begin{center}
\includegraphics[width=1.0\textwidth]{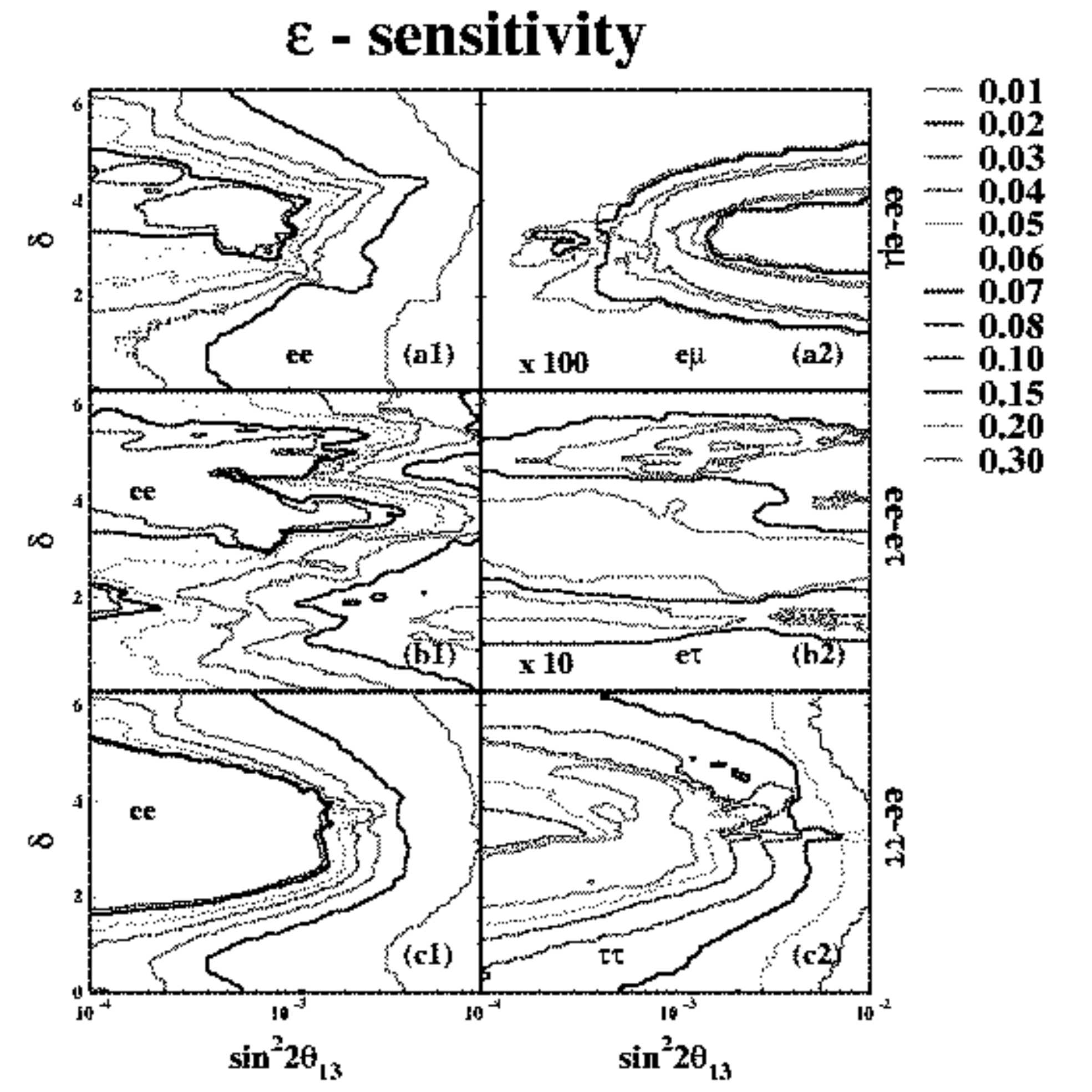}
\end{center}
\vglue -0.6cm
 \caption{
Iso-contours of 2 $\sigma$ CL (for 2 DOF) 
sensitivity (uncertainty) for $\varepsilon_{\alpha\beta}$ 
parameter as functions of the input parameters of
$\sin^2 2\theta_{13}$  and $\delta$.
For each point in the $\sin^2 2\theta_{13}-\delta$ plane
the uncertainty is defined as 
$\Delta \varepsilon \equiv 
(\varepsilon_{\text{max}}-\varepsilon_{\text{min}})/2$ 
where $\varepsilon_{\text{max(min)}}$ indicates 
the maximum and minimum allowed value of 
$\varepsilon$ parameters which 
is consistent with the case without NSI effect. 
In the upper, lower and bottom panels, the sensitivities for 
(a1) $\varepsilon_{ee}$ and (a2) $\varepsilon_{e\mu}$ 
for the $\varepsilon_{ee}$-$\varepsilon_{e\mu}$ system, 
(b1) $\varepsilon_{ee}$  and  (b2) $\varepsilon_{e\tau}$
for the $\varepsilon_{ee}$-$\varepsilon_{e\tau}$ system, 
and 
(c1) $\varepsilon_{ee}$ and  (c2) $\varepsilon_{\tau\tau}$
for the $\varepsilon_{ee}$-$\varepsilon_{\tau\tau}$ system, 
respectively, are shown. 
We note that the 
the uncertainty for $\varepsilon_{e\mu}$ shown in (a2)
and $\varepsilon_{e\tau}$ shown in (b2) are 
magnified by 100 and 10, respectively. 
Some of the structures are due to the finite grid used in our 
calculation, so the precise shape of the contours may not be reliable. 
}
\label{epsilon-sens1}
\end{figure}

\begin{figure}[htbp]
\begin{center}
\includegraphics[width=1.0\textwidth]{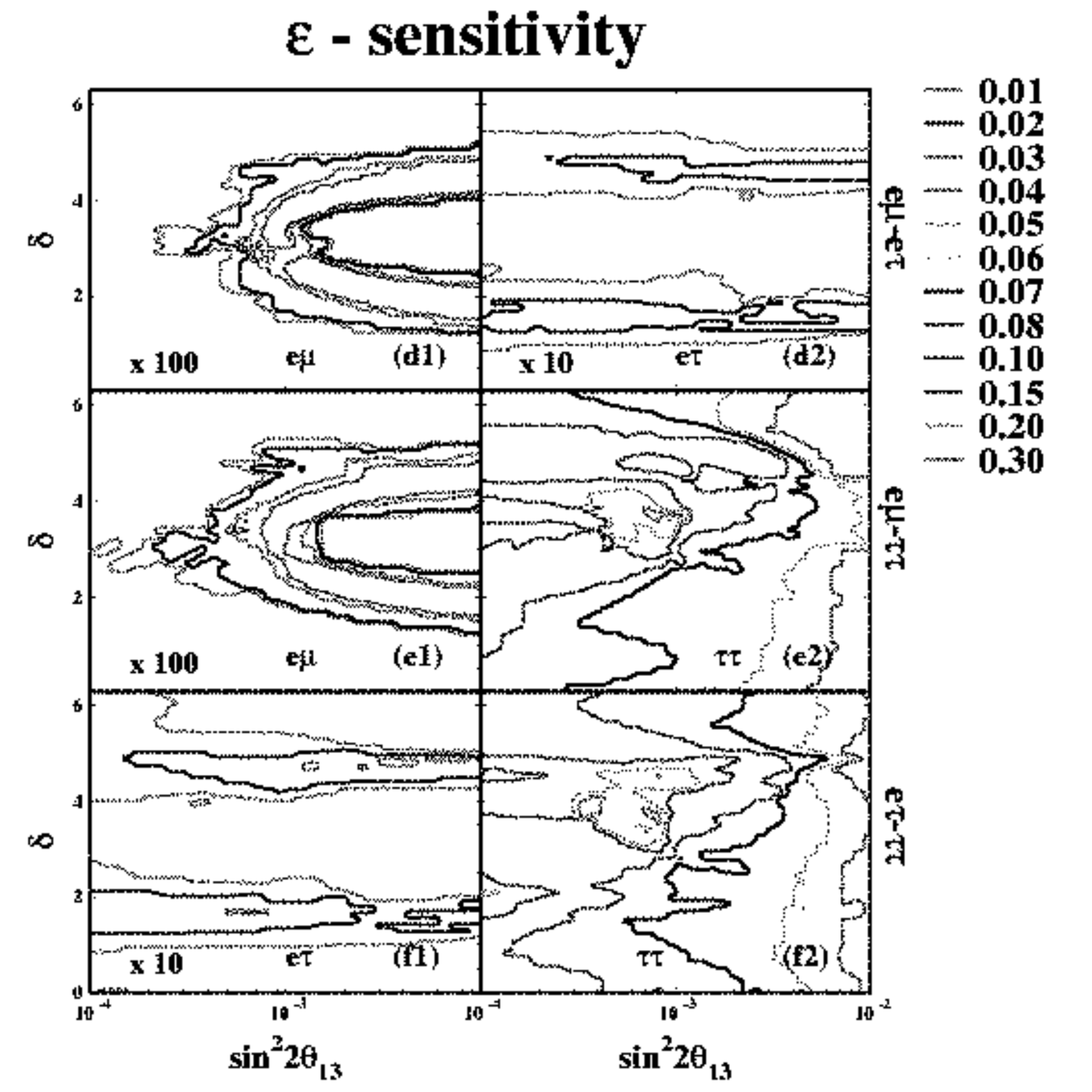}
\end{center}
\vglue -0.6cm
\caption{
Same as in Fig.~\ref{epsilon-sens1} but for 
different combination of the 2 $\varepsilon$ system. 
In the upper, lower and bottom panels, the sensitivities for 
(d1) $\varepsilon_{e\mu}$ and (d2) $\varepsilon_{e\tau}$ 
for the $\varepsilon_{e\mu}$-$\varepsilon_{e\tau}$ system, 
(e1) $\varepsilon_{e\mu}$ and  (e2) $\varepsilon_{\tau\tau}$
for the $\varepsilon_{e\mu}$-$\varepsilon_{\tau\tau}$ system, 
and 
(f1) $\varepsilon_{e\tau}$ and (f2) $\varepsilon_{\tau\tau}$ 
for the $\varepsilon_{e\tau}$-$\varepsilon_{\tau\tau}$ system, 
respectively, are shown. 
As in Fig.~\ref{epsilon-sens1}
the uncertainty for $\varepsilon_{e\mu}$ shown in (d1) and (e1)
and $\varepsilon_{e\tau}$ shown in (d2) and (f1) are 
magnified by 100 and 10, respectively. 
}
\label{epsilon-sens2}
\end{figure}

\begin{figure}[htbp]
\begin{center}
\includegraphics[width=0.9\textwidth]{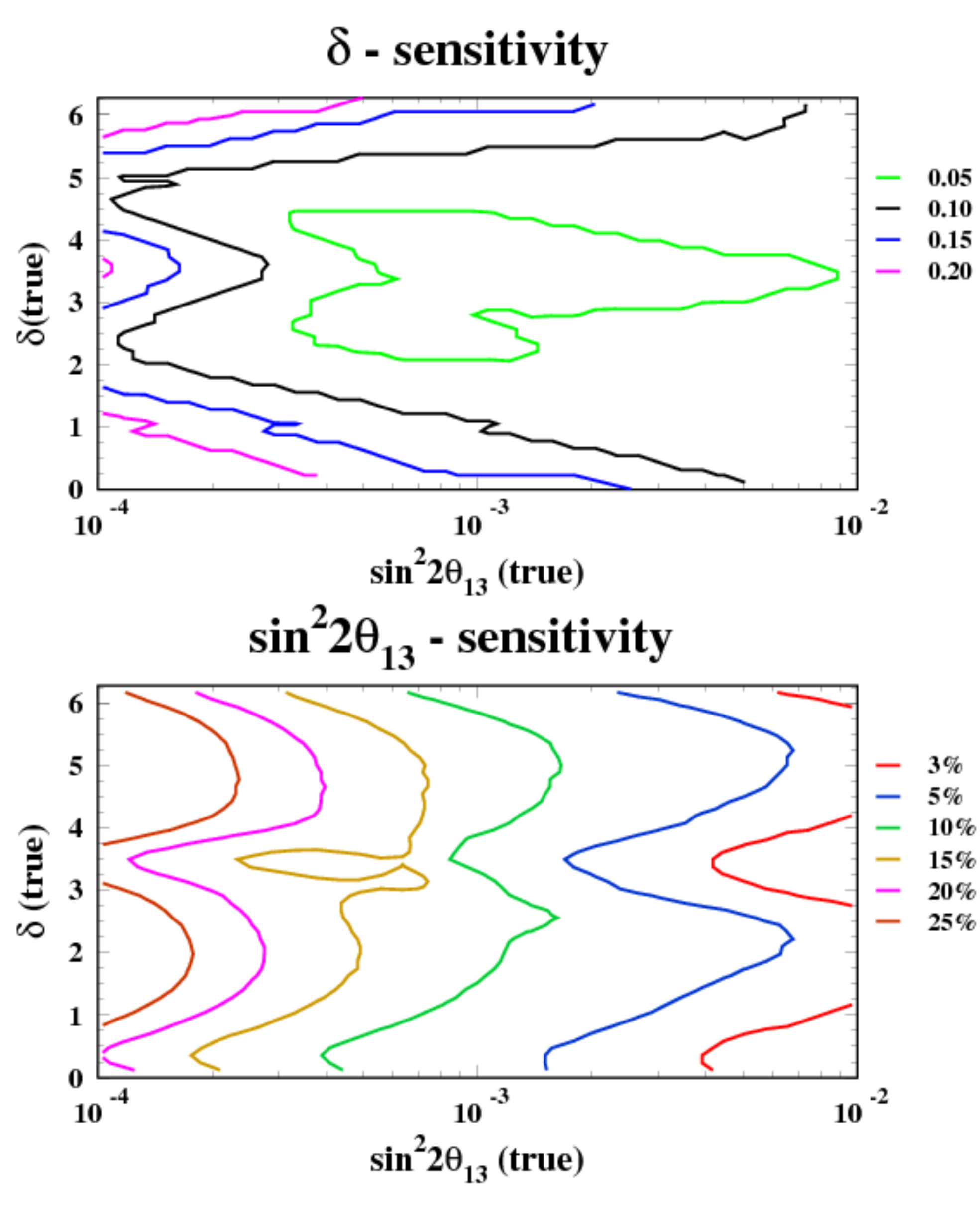}
\end{center}
\vglue -1.0cm
\caption{
Upper panel: Iso-contours of 2 $\sigma$ CL (for 2 DOF)  
sensitivity (uncertainty) for the CP phase $\delta$ (in radians) 
expected to be achieved at neutrino factory in the absence of 
the NSI effect in the plane of 
the true values of $\delta$ and $\sin^2 2\theta_{13}$. 
The uncertainty is defined as 
$\Delta \delta \equiv 
(\delta_{\text{max}}-\delta_{\text{min}})/2$ 
in radians, where 
$\delta_{\text{max(min)}}$ is maximum (minimum) allowed 
value of $\delta$ (mod. $2\pi$) for each given input point. 
Lower panel: Similar plot as in the upper panel but for the fractional
uncertainty $\Delta(\sin^2 2\theta_{13})/\sin^2 2\theta_{13}$ (in
percent) is shown.  }
\label{sens-std}
\end{figure}

\begin{figure}[htbp]
\begin{center}
\includegraphics[width=0.9\textwidth]{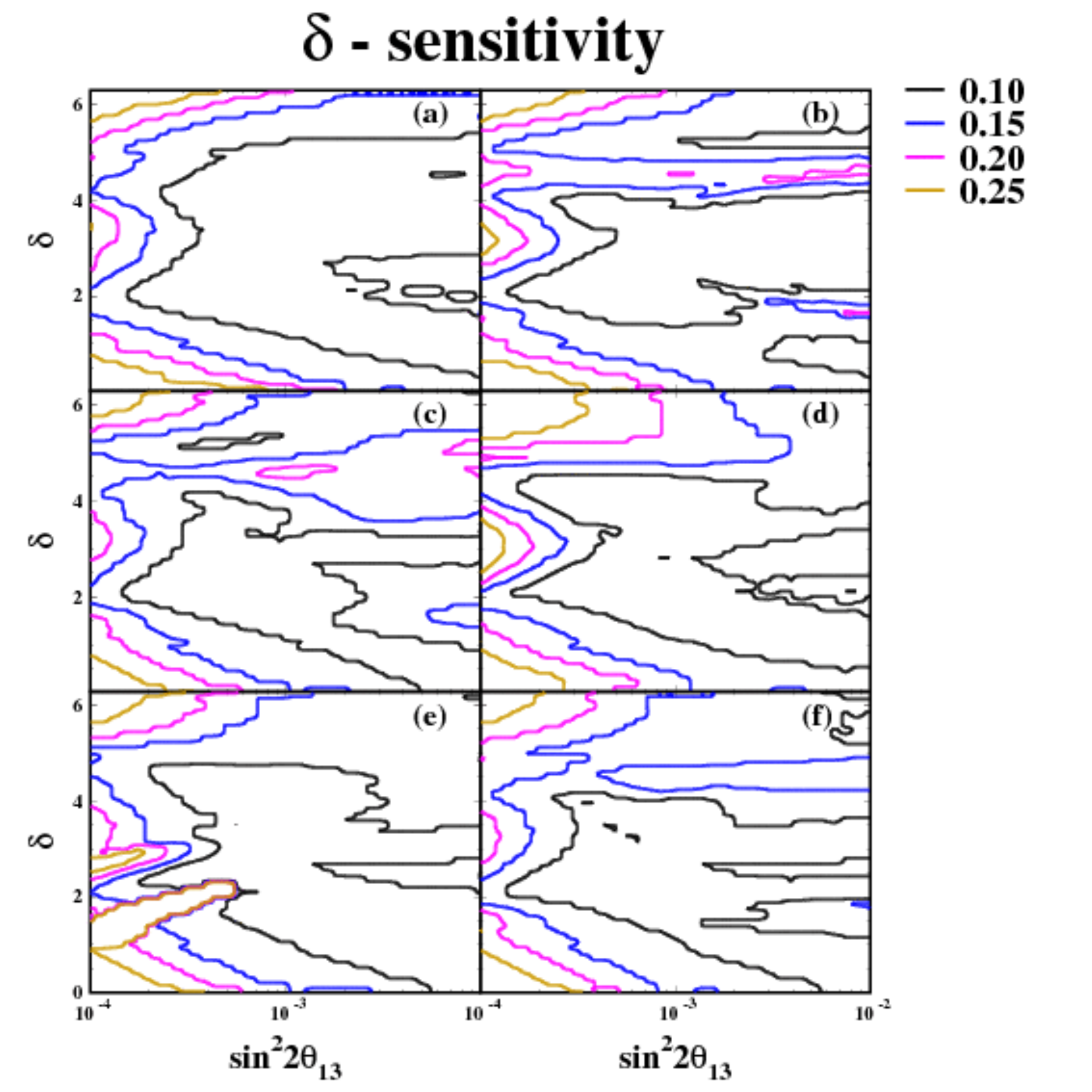}
\end{center}
\vglue -0.6cm
\caption{
Same as in the upper panel of Fig.~\ref{sens-std} but
for the case where the NSI parameters are turned on
in the fit.
The iso-contours of $\Delta \delta$ (in radians) 
at 2 $\sigma$ CL (for 2 DOF) 
are shown for the 6 combinations 
of 2 $\varepsilon$ system: 
(a) $\varepsilon_{ee}-\varepsilon_{e\mu}$, 
(b) $\varepsilon_{ee}-\varepsilon_{e\tau}$, 
(c) $\varepsilon_{ee}-\varepsilon_{\tau\tau}$, 
(d) $\varepsilon_{e\mu}-\varepsilon_{e\tau}$,  
(e) $\varepsilon_{e\mu}-\varepsilon_{\tau\tau}$ and
(f) $\varepsilon_{e\tau}-\varepsilon_{\tau\tau}$. 
}
\label{del-sens}
\end{figure}

\begin{figure}[htbp]
\begin{center}
\includegraphics[width=0.9\textwidth]{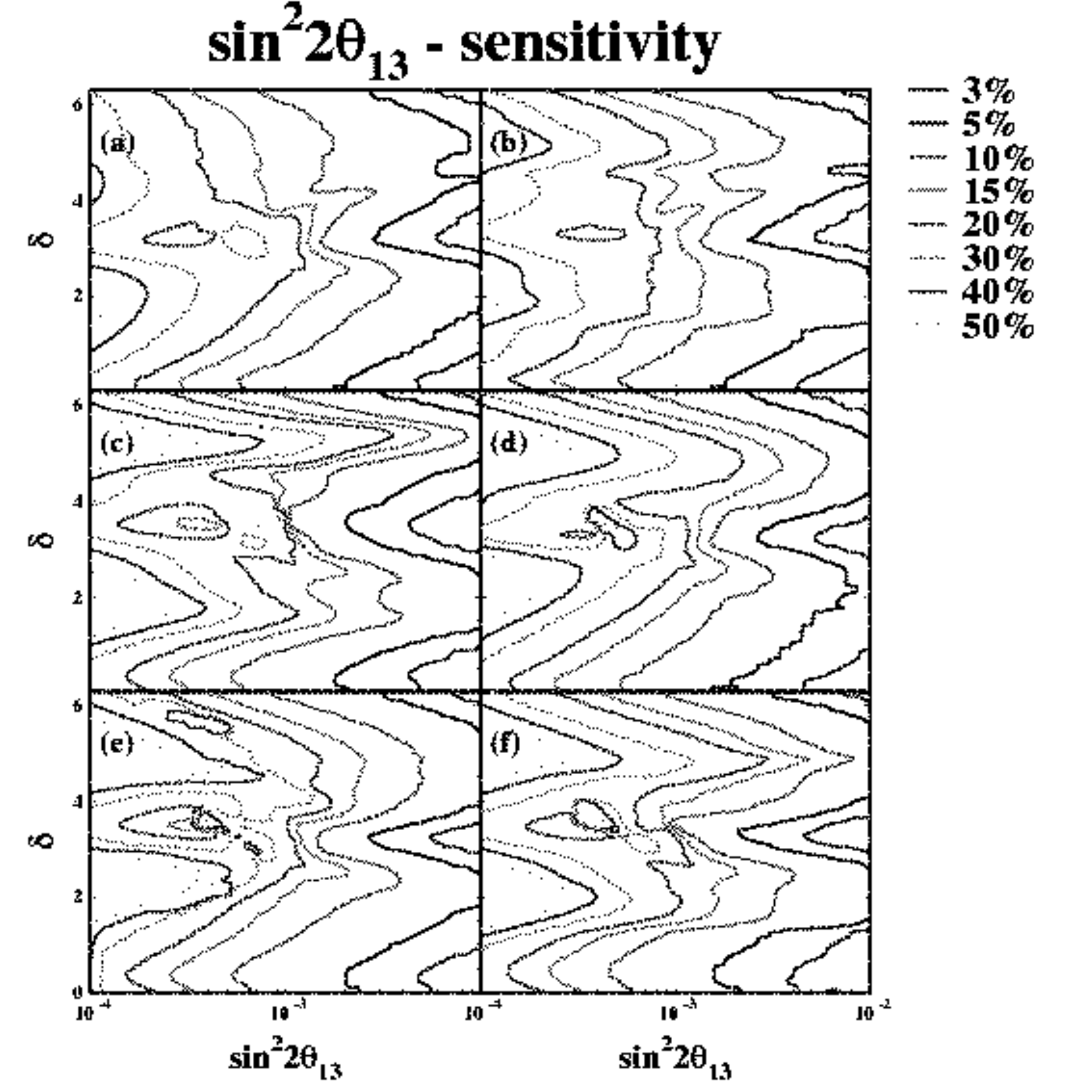}
\end{center}
\caption{
Same as in the lower panel of Fig.~\ref{sens-std} but
for the case where the NSI parameters are turned on
in the fit.
The iso-contours of 
$\Delta(\sin^2 2\theta_{13})/\sin^2 2\theta_{13}$ at 
2 $\sigma$ CL (for 2 DOF) 
are shown for the 6 combinations 
of 2 $\varepsilon$ system:
(a) $\varepsilon_{ee}-\varepsilon_{e\mu}$, 
(b) $\varepsilon_{ee}-\varepsilon_{e\tau}$, 
(c) $\varepsilon_{ee}-\varepsilon_{\tau\tau}$, 
(d) $\varepsilon_{e\mu}-\varepsilon_{e\tau}$,  
(e) $\varepsilon_{e\mu}-\varepsilon_{\tau\tau}$ and
(f) $\varepsilon_{e\tau}-\varepsilon_{\tau\tau}$. 
}
\label{th13-sens}
\end{figure}

In Figs.~\ref{epsilon-sens1}-\ref{th13-sens} we have presented 
the equi-uncertainty contours of a particular observable 
$O$ in the plane spanned by the input values of 
$\sin^2 2\theta_{13}$ and $\delta$. 
We have defined the uncertainty of measuring (or constraining) the 
observable $O$ as  $\Delta O \equiv (O_{\rm max} -O_{\rm min})/2$, 
except for the case of $\sin^2 2\theta_{13}$, where we give the fractional 
uncertainty, i.e. $\Delta (\sin^2 2\theta_{13})/\sin^2 2\theta_{13}$.
For all cases we present the 2$\sigma$ CL (2 DOF) contours.

The sensitivity to the NSI parameters, $\theta_{13}$ and $\delta$ 
can be determined in the analysis of the 6 combinations of NSI 
elements we have examined in this work:
$\varepsilon_{e e} - \varepsilon_{e \mu}$, 
$\varepsilon_{e e} - \varepsilon_{e \tau}$,
$\varepsilon_{e e} - \varepsilon_{\tau \tau}$,
$\varepsilon_{e \mu} - \varepsilon_{e \tau}$, 
$\varepsilon_{e \mu} - \varepsilon_{\tau \tau}$ and 
$\varepsilon_{e \tau} - \varepsilon_{\tau \tau}$. 
Therefore we need altogether 12 panels to fully present the sensitivities 
to the $\varepsilon$'s, and 12 panels present the sensitivities to 
the standard oscillation parameters $\theta_{13}$ (6 panels) and $\delta$ 
(6 panels). 

We show, for definiteness, the case the NSI parameters  have 
zero (vanishingly small) input values.  But, as we saw in 
Sec.~\ref{nonzero-input}, the results would not be so different even 
if we had taken non-zero input values for the NSI. 
Prior to showing the sensitivity contours we wish to warn the readers; 
Some of the structures are due to the finite grid used in our 
calculation, so the precise shape of the contours may not be reliable.

Let us first look at Figs.~\ref{epsilon-sens1} and \ref{epsilon-sens2} 
where we show the sensitivity to the NSI parameters.  
Roughly speaking, for the typical value $\sin^2 2\theta_{13} \sim 10^{-3}$ 
the sensitivity to $\varepsilon_{\tau \tau}, \varepsilon_{e e}, 
\varepsilon_{e \tau}$ and $\varepsilon_{e \mu}$ are 
$\sim$ 10-20\%,  2-10\%,  0.1-0.4\% and 0.01-0.04\%, 
respectively.  
As we have discussed in previous sections,  the off-diagonal NSI parameters 
have a much more significant impact than that of the diagonal ones. We 
confirm here that in fact, they can be significantly more constrained 
by data. 
We observe that the sensitivity to the off-diagonal NSI  parameters is 
basically not affected by the presence of another non-zero NSI contribution, 
whereas this is not true for the sensitivity to the diagonal elements. 
We suspect that this comportment may continue to be true even if more NSI 
parameters are switched on simultaneously.

Some of the features of these contours can be readily understood from 
the bi-probability plots in Fig.~\ref{bi-Pmagic}.
For example, from panel (b2) of Fig.~\ref{epsilon-sens1} and panels  
(d2) and (f1) of Fig.~\ref{epsilon-sens2}, we observe that 
the sensitivity to $\varepsilon_{e\tau}$ is best at $\delta=0$ and 
$\delta=\pi$ and worst at $\delta=\pi/2$ and $3\pi/2$. 
This is exactly what one would expect from the upper right panel of  
Fig.~\ref{bi-Pmagic}, namely, the points in $P-\bar{P}$ space 
corresponding to $\delta=\pi/2$ (square) and $\delta=3\pi/2$ (diamond)
with non-zero $\varepsilon_{e\tau}$  can be confused with that of the 
standard case without NSI effect (orange strip). 
One can also understand that this behavior does not depend on $\theta_{13}$. 

Despite the contours for $\varepsilon_{e e}$ have a more
complicated structure, one can guess from the upper left panel of
Fig.~\ref{bi-Pmagic} that the best sensitivity should be for $\delta
\sim 0-\pi/4$ and worse for $\delta \sim \pi$. This is also confirmed 
by panels (a1), (b1) and (c1) of Fig.~\ref{epsilon-sens1}.

The sensitivity to $\varepsilon_{e\mu}$ shows an intricate dependence 
on $\delta$ and $\theta_{13}$ as seen in Fig.~\ref{epsilon-sens1}(a2) and
Figs.~\ref{epsilon-sens2}(d1) and (e1). 
The sensitivity is the worst at around 
$\delta \simeq \pi/2$ and $\simeq 3\pi/2$ for $\sin^2 2\theta_{13} \sim 10^{-2}$, 
and at around 
$\delta \simeq \pi$ for $\sin^2 2\theta_{13} \sim 10^{-3}$. 
At $\sin^2 2\theta_{13} \sim 10^{-4}$, the sensitivity to 
$\varepsilon_{e\mu}$ has no significant dependence on $\delta$. 
Let us understand the cause of such complicated features.  
First of all, the sensitivity to $\varepsilon_{e\mu}$ is coming dominantly 
from the measurement at 3000 km, as indicted in 
Figs.~\ref{ee-em-mm-piby4}, \ref{ee-em-et-3piby2} and 
\ref{ee-em-mm-3piby2-0.0001}. 
Then, it should be possible to understand the behavior at 
$\sin^2 2\theta_{13} \sim 10^{-3}$ from the bi-probability plot 
give in Fig.\ref{bi-P3000}. 
This is indeed the case; The upper middle panel of the bi-probability 
plot shows that the ellipses of positive and negative 
$\varepsilon_{e\mu}$ as well as the one without NSI 
overlap with each other at $\delta \sim \pi$, 
and hence the sensitivity is worst at $\delta \sim \pi$. 
The similar bi-probability plot for $\sin^2 2\theta_{13} \sim 10^{-2}$  
(not shown)  indicates that a positive, negative, and zero  
$\varepsilon_{e\mu}$ ellipses meet at $\delta \sim \pi/2$ and 
$\sim 3\pi/2$. 
(Precise values of the crossing point are at around 
$\delta = 0.6\pi$ and $1.4\pi$.) 
At $\sin^2 2\theta_{13} \sim 10^{-4}$ (not shown) 
there is no particular value of $\delta$ at which these ellipses overlap. 
Thus, these features of the bi-probability plots explain the behavior 
of sensitivity contours of $\varepsilon_{e\mu}$ in 
Fig.~\ref{epsilon-sens1}(a2) and
Figs.~\ref{epsilon-sens2}(d1) and (e1).

Let us also discuss the sensitivity reach of $\theta_{13}$ and
$\delta$ with and without NSI effect.  In Fig.~\ref{sens-std} we show
the iso-contours of uncertainty on the determination of the 
CP phase $\delta$ (upper panel) and of $\sin^2 2\theta_{13}$ (lower panel) 
in the plane spanned by the input (true) values of 
$\sin^2 2\theta_{13}$ and $\delta$ in the absence of NSI effect. 
These results must be compared with the case with NSI effect we will 
show below.

In Fig.~\ref{del-sens} we show the iso-uncertainty contour of the CP
phase $\delta$ in the presence of NSI parameters for the 6 different
combinations of  the 2 $\varepsilon$ systems we considered in this work.
Roughly speaking, for the typical input value $\sin^2 2\theta_{13}
\sim 10^{-3}$ the sensitive become worse when NSI is included.
Nevertheless, the change is from $\Delta \delta \simeq 0.05-0.1$ radians 
(without NSI) to $\simeq 0.1-0.15$ radians or so (with NSI), in all 6 cases.  
We conclude that the difference between the sensitivities to $\delta$ 
with and without NSI is not dramatic.

On the other hand, regarding the sensitivity to $\sin^2
2\theta_{13}$, comparing the lower panel of Fig.~\ref{sens-std} with 
Fig.~\ref{th13-sens}, we can see that for the typical input value
$\sin^2 2\theta_{13} \sim 10^{-3}$ the fractional uncertainty on
$\sin^2 2\theta_{13}$ becomes larger with NSI, going 
from $\Delta(\sin^2 2\theta_{13})/\sin^2 2\theta_{13}\sim 10$\% (without NSI)
to $\sim 10-20$\% (with NSI).  Again we conclude that the impact of NSI 
in the determination of $\theta_{13}$ is not so striking.

\section{Concluding remarks} 
\label{conclusion} 

We have demonstrated in this paper that a neutrino factory equipped with 
an intense neutrino flux from a muon storage ring and two detectors, one 
located  at $L=3000$ km and the other at $L=7000$ km, is powerful enough 
to probe into extremely small values of the NSI parameters. 
We have relied on the golden channel, 
$\nu_{e} \rightarrow \nu_{\mu}$, and its anti-neutrino counter part 
in our analysis in this paper. 
Six different combinations of the two $\varepsilon$ systems that
can be obtained from the four NSI parameters,
$\varepsilon_{\alpha\beta}$ with $(\alpha,\beta)$ being $(e,e)$,
$(\tau, \tau)$, $(e,\mu)$ and $(e, \tau)$, are analyzed under the
assumption of ignoring the effects of NSI at the production and the
detection of neutrinos.

The sensitivities to off-diagonal $\varepsilon$'s are excellent, 
$|\varepsilon_{e \tau}| \simeq \text{a few} \times10^{-3}$ and 
$|\varepsilon_{e \mu}| \simeq \text{a few} \times10^{-4}$ and 
while the ones for the diagonal $\varepsilon$'s are acceptable, 
$|\varepsilon_{e e}| (|\varepsilon_{\tau \tau}|) \simeq 0.1 (0.2)$ 
at 3$\sigma$ CL and 2 DOF. 
These sensitivities remain more or less independent of $\theta_{13}$ 
down to extremely small values such as $\sin^2 2\theta_{13}=10^{-4}$. 
They seem also very robust in the sense that they are 
not very disturbed by the presence of another non-zero NSI 
contribution.
The above characteristics of the sensitivities to NSI suggest that 
in our setting the off-diagonal $\varepsilon$'s are likely the best 
place to discover NSI. 
We note that these results are obtained under the 
assumption of ignoring background as well as systematic errors 
in our analysis. 
%
According to our estimate, however, the effect of inclusion of them 
is limited to $\sim 50$\% (a factor of 2) 
or so for $\theta_{13}$ as small as $\sin^2 2\theta_{13} = 10^{-3}$ 
($10^{-4}$).

One of the most significant features of the results obtained in our
analysis is that the presence of NSI {\em does not} confuses the
precision measurement of $\theta_{13}$ and $\delta$.  The favorite
property arises from the synergy between the two detectors placed at
the two different baselines;
The detector at the magic baseline is extremely (reasonably) sensitive
to the off-diagonal (diagonal) $\varepsilon$'s but lack sensitivity to
$\delta$.  On the other hand, the intermediate detector at $L=3000$ km
{\em is} sensitive to $\delta$, while lacking good sensitivity to
$\varepsilon$'s (except in the case of $\varepsilon_{e \mu}$).  
We have shown in Sec.~\ref{zero-input} and
\ref{nonzero-input} that, when combined, the synergy between the two
detectors has an enormous power to resolve the confusion between
$\theta_{13}$ and NSI.
Moreover, the impact of the systematic errors on $\delta$ and 
$\theta_{13}$ is limited especially at small $\theta_{13}$ and 
at most $\sim$10\%-20\% level at $\sin^2 2\theta_{13} = 10^{-4}$. 
We believe that the results obtained in this paper open the door to
the possibility of using neutrino factory as a discovery machine for
NSI.

Our analysis in this paper, however, has limitations of validity
because we have assumed that the effects of NSI in production and the
detection processes are negligible.  It may be a self-consistent
approximation when we discuss the $\varepsilon_{e \tau}$ system, assuming 
a very small $\varepsilon_{e \tau}$, because its effect on muon decay is
expected to be small.  But, if we discuss the $\varepsilon_{e \mu}$ system 
it affects the production and the detection processes at one-loop
level and all these effects have to be consistently dealt with.  In this
paper we are not able to address this point.
Our analysis also does not include the energy resolution and the
uncertainty on the matter density as well as on the remaining mixing
parameters.

We have also mentioned in Sec.~\ref{NSI} that the phase of the 
off-diagonal $\varepsilon_{\alpha \beta}$ produces confusion problems of
two types, the two-phase and the phase-magnitude confusion.
In our analysis we have shown that the discrete version of the former
degeneracy is resolved by the two-detector setting examined in this
paper.  Therefore, it is natural to expect that the degeneracies in
its generic form as well as the latter type will also be resolved by
the two-detector setting.  The magnitudes of the solar 
$\Delta m^2_{21}$ sensitive terms will be different between the intermediate
and the far detectors so that the confusions will be lifted by
combining them.  The work toward this direction is in progress.

One could think about adding the silver channel in the analysis
\cite{campanelli}.  This could, in principle, increase the sensitivity
to $\epsilon_{\tau x}$.  In the present setting of magnetized iron
detectors it requires separation of muons produced by tau decay from
those of oscillated $\nu_{\mu}$ CC reaction.  This requires more
sophisticated analysis and therefore it is left for future
investigation.

\begin{acknowledgments}
  Four of us (H.M., H.N., S.U. and R.Z.F) are grateful for the
  hospitality of the Theory Group of the Fermi National Accelerator
  Laboratory during the summer of 2007 where this work was completed.
  H.M. and R.Z.F thank Departamento de F\'{\i}sica, Pontif{\'\i}cia
  Universidade Cat{\'o}lica do Rio de Janeiro, where part of this work
  was carried out, for the hospitality.
  This work was supported in part by KAKENHI, Grant-in-Aid for
  Scientific Research, No 19340062, Japan Society for the Promotion of
  Science, by Funda\c{c}\~ao de Amparo \`a Pesquisa do Estado de S\~ao
  Paulo (FAPESP), Funda\c{c}\~ao de Amparo \`a Pesquisa do Estado de
  Rio de Janeiro (FAPERJ) and Conselho Nacional de Ci\^encia e
  Tecnologia (CNPq). 
\end{acknowledgments}

\clearpage

\clearpage

\clearpage


\newpage

\appendix

\section{Derivation of appearance probability 
$P(\nu_{e} \rightarrow \nu_{\mu})$ in the presence of 
non-standard interactions} 
\label{derivation}

In this Appendix, we give a self-contained discussion for deriving the
expression of the appearance probability $P(\nu_{e} \rightarrow
\nu_{\mu})$ with simultaneous presence of non-standard interactions
$\varepsilon_{e e}$ and $\varepsilon_{e \tau}$.  For simplicity, we
denote it $\varepsilon_{e \tau} - \varepsilon_{e e}$ system, and use
the method developed by Kimura, Takamura, and Yokomakura
(KTY)~\cite{KTY}. 
(See \cite{blom} for minor sign error in the original 
formula, and \cite{yasuda} for a reformulation.) 

The evolution equation of neutrinos can be written in the 
flavor eigenstate as 
\begin{eqnarray}
i\frac{d}{dx} \nu_{\alpha} = 
\frac{1}{2E} H_{\alpha \beta} \nu_{\beta}
\hspace*{15mm}
(\alpha, \beta = e, \mu, \tau),
\label{evolution}
\end{eqnarray}
where the Hamiltonian is given by
\begin{eqnarray}
H=
U \left[
\begin{array}{ccc}
\Delta m^2_{11} & 0 & 0 \\
0 & \Delta m^2_{21}& 0 \\
0 & 0 & \Delta m^2_{31} 
\end{array}
\right] U^{\dagger}
+
a(x) \left[
\begin{array}{ccc}
1 + \varepsilon_{e e} & 0 & \varepsilon_{e \tau} \\
0 & 0 & 0 \\
\varepsilon_{e \tau}^{*} & 0 & 0
\end{array}
\right], 
\label{hamiltonian}
\end{eqnarray}
whose first term will be denoted as $H^{\rm vac}$ hereafter 
and $\Delta m^2_{ji} \equiv m^2_{j} - m^2_{i}$. 
(Hence, $\Delta m^2_{11} \equiv 0$ by definition.)
In (\ref{hamiltonian}), $a \equiv 2\sqrt{2} G_F n_e(x) E$ denotes 
the coefficient related to the index of refraction of neutrinos in medium 
\cite{wolfenstein} with electron number density $n_e(x)$, where 
$G_F$ is the Fermi constant and $E$ is the neutrino energy. 
Despite that $n_e(x)$ may depend upon locations along 
the neutrino trajectory, we use constant density 
approximation throughout this paper. 
The MNS matrix $U$ relates the flavor and the vacuum 
mass eigenstates as 
\begin{eqnarray}
\nu_{\alpha} = (U)_{\alpha i} \nu_{i} ,
\end{eqnarray}
where $i$ runs over 1-3. 
We use the standard parametrization of the MNS matrix \cite{PDG}; 
\begin{eqnarray}
U=\left[
\begin{array}{ccc}
c_{12}c_{13} & s_{12}c_{13} &   s_{13}e^{-i\delta}\\
-s_{12}c_{23}-c_{12}s_{23}s_{13}e^{i\delta} &
c_{12}c_{23}-s_{12}s_{23}s_{13}e^{i\delta} & s_{23}c_{13}\\
s_{12}s_{23}-c_{12}c_{23}s_{13}e^{i\delta} &
-c_{12}s_{23}-s_{12}c_{23}s_{13}e^{i\delta} & c_{23}c_{13}\\
\end{array}
\right],
\label{MNSmatrix}
\end{eqnarray}
where  $c_{ij}$ and $s_{ij}$ ($i, j = 1 \mbox{-} 3$) imply 
$\cos{\theta_{ij}}$ and $\sin{\theta_{ij}}$,  respectively.

By defining renormalized matter coefficient and 
renormalized $\varepsilon_{e \tau}$ as 
\begin{eqnarray}
\tilde{a} \equiv  a ( 1 + \varepsilon_{e e} ) 
\hspace{20mm}
\tilde{ \varepsilon}_{e \tau} \equiv  \frac{ \varepsilon_{e \tau} }{ 1 + \varepsilon_{e e}  }  
\label{renormalized}
\end{eqnarray}
the matter term in the Hamiltonian can be written as 
\begin{eqnarray}
\tilde{a}  \left[
\begin{array}{ccc}
1 & 0 & \tilde{ \varepsilon}_{e \tau}  \\
0 & 0 & 0 \\
\tilde{ \varepsilon}_{e \tau}^{*} & 0 & 0
\end{array}
\right]. 
\label{matter} 
\end{eqnarray}
Thus, the problem is reduced to the effective system with only 
single type of NSI, $\tilde{ \varepsilon}_{e \tau}$. 

We now define the mass eigenstate in matter $\nu_{i}^{m}$ 
by using the transformation 
\begin{eqnarray}
\nu_{\alpha} = (V)_{\alpha i} \, \nu_{i}^{m},
\label{defV}
\end{eqnarray}
where $V$ is the unitary matrix which diagonalize the Hamiltonian 
with scaled eigenvalues $\lambda$ as 
$V^{\dagger} H V = H_{\rm diag} 
\equiv diag(a\lambda_1, a\lambda_2, a\lambda_3)$. 
We first obtain the expressions of the eigenvalues of the 
Hamiltonian (\ref{hamiltonian}). 
They are determined by the equation 
$det[H - \lambda a I]=0$ which is the cubic equation 
for the scaled eigenvalue $\lambda$:
\begin{eqnarray}
&&\lambda^3 -  
(1 + \delta_{31} + \delta_{21}) \lambda^2 
\nonumber \\ 
&+& 
\Bigl[ 
c_{13}^2 \delta_{31} + 
\left \{ 
\delta_{31} + c_{12}^2  + s_{12}^2 s_{13}^2 + 
2 c_{12} s_{12} s_{23} c_{13} Re(\tilde{ \varepsilon}_{e \tau})  
\right \} \delta_{21} 
\Bigr.
\nonumber \\
& & 
\Bigl.
\hspace*{46mm}
- 2 c_{23} c_{13} s_{13} (\delta_{31} - s^2_{12} \delta_{21}) Re(\tilde{ \varepsilon}_{e \tau} e^{i \delta}) - \vert \tilde{ \varepsilon}_{e \tau} \vert^{2}
\Bigr] \lambda
\nonumber \\ 
&-& 
\delta_{21} \delta_{31} 
\Bigl[ 
c_{12}^2 c_{13}^2 + 2 c_{12} s_{12} s_{23} c_{13} Re(\tilde{ \varepsilon}_{e \tau}) - 
2  c^2_{12} c_{23} c_{13} s_{13} Re(\tilde{ \varepsilon}_{e \tau} e^{i \delta}) 
\Bigr] 
\nonumber \\
&+&
\vert \tilde{ \varepsilon}_{e \tau} \vert^{2} 
\Bigl[ 
s^2_{23} c^2_{13} \delta_{31} + 
( c_{12}^2 c_{23}^2 +  s_{12}^2 s_{23}^2 s_{13}^2 - 2 c_{12} s_{12} c_{23} s_{23} s_{13} \cos \delta  ) \delta_{21} 
\Bigr]
= 0. 
\label{eigenvalue-eq}
\end{eqnarray}
in which everything is scaled by $a$ and 
$\delta_{21}$ and $\delta_{31}$ denote the scaled squared 
mass differences,
\begin{eqnarray}
\delta_{21} \equiv \frac{\Delta m^2_{21}}{\tilde{a}}, 
\hspace*{15mm}
\delta_{31} \equiv \frac{\Delta m^2_{31}}{\tilde{a}}.
\end{eqnarray} 

\subsection{KTY method for obtaining exact oscillation probability with NSI}

We follow the KTY method \cite{KTY} for deriving 
$P(\nu_{e} \rightarrow \nu_{\mu})$ and write down the equations 
\begin{eqnarray}
H_{e \mu} &=& H^{vac}_{e \mu},  
\nonumber \\
H_{e \tau} H_{\tau \mu} - H_{e \mu} H_{\tau \tau} &=&
(H^{vac}_{e \tau}  + \tilde{ \varepsilon}_{e \tau} ) H^{vac}_{\tau \mu} - 
H^{vac}_{e \mu} H^{vac}_{\tau \tau}. 
\label{KTYeq1}
\end{eqnarray}
They give relationships between mixing matrix in vacuum and in matter 
as 
\begin{eqnarray}
\sum_{i} \lambda_{i} V_{e i} V_{\mu i}^*  &=&
\sum_{i} \delta_{j1}  U_{e i} U_{\mu i}^*  
\equiv p, 
\nonumber \\
\sum_{ijk}^{cyclic} \lambda_{j}\lambda_{k} V_{e i} V_{\mu i}^*  &=& 
\sum_{ijk}^{cyclic} \delta_{j1} \delta_{k1} U_{e i} U_{\mu i}^* + 
\tilde{ \varepsilon}_{e \tau} \sum_{i} \delta_{i1} U_{\tau i} U_{\mu i}^* 
\equiv q.
\label{KTYeq2}
\end{eqnarray}
Notice that the effect of $\tilde{ \varepsilon}_{e \tau} $ is contained only in $q$. 
Solving (\ref{KTYeq2}) for $V_{e i} V_{\mu i}^*$ 
under the constraint of unitarity 
$\sum_{i} V_{e i} V_{\mu i}^*  = 0 $ 
we obtain 
\begin{eqnarray}
 V_{e i} V_{\mu i}^* = \frac{p \lambda_{i} + q}
{ \Delta_{ji} \Delta_{ki} } 
\label{VV} 
\end{eqnarray}
where $\Delta_{ji} \equiv \lambda_{j} - \lambda_{i}$ and 
($i,j,k$) are cyclic.

Then, the appearance probability $P(\nu_{e} \rightarrow \nu_{e})$ 
is given exactly by  \cite{KTY},  
\begin{eqnarray}
P(\nu_e \to \nu_{\mu}) &=& 
4 \sum_{(ijk)}^{{\rm cyclic}} 
({\rm Re}\tilde{J}_{e\mu}^{ij}  + {\rm Re}\tilde{J}_{e\mu}^{jk} )
\cos \left(\frac{ \tilde{a} L}{4E} \Delta_{ki} \right) 
\sin \left(\frac{ \tilde{a} L}{4E} \Delta_{ij} \right) 
\sin \left(\frac{ \tilde{a} L}{4E} \Delta_{jk} \right), 
\nonumber \\
&+& 
8 \sum_{(ijk)}^{{\rm cyclic}} 
\tilde{J} 
\sin \left(\frac{ \tilde{a} L}{4E} \Delta_{12} \right) 
\sin \left(\frac{ \tilde{a} L}{4E} \Delta_{23} \right) 
\sin \left(\frac{ \tilde{a} L}{4E} \Delta_{31} \right), 
\label{Pemu_exact}
\end{eqnarray}
where the sum over the cyclic permutation is implied and  
\begin{eqnarray}
{\rm Re}\tilde{J}_{e\mu}^{ij} &=&
\frac{|p|^2 \lambda_i \lambda_j+|q|^2
+{\rm Re}(pq^*)(\lambda_i+\lambda_j)}
{\Delta_{ij}\Delta_{12}\Delta_{23}
\Delta_{31}}, 
\label{6} \\
\tilde{J}&=&\frac{{\rm Im}(pq^*)}
{\Delta_{12}\Delta_{23}
\Delta_{31}}. 
\label{Jemuij}
\end{eqnarray}
It is thus convenient to compute the combination 
${\rm Re} \tilde{J}_{e\mu}^{ij} + {\rm Re} \tilde{J}_{e\mu}^{jk} $ 
(note the minus sign), 
\begin{eqnarray}
{\rm Re} \tilde{J}_{e\mu}^{ij} + {\rm Re} \tilde{J}_{e\mu}^{jk} 
\equiv \frac{ - 1 }{ ( \Delta_{ij} \Delta_{jk} )^2  } J_{j} 
\end{eqnarray}
where 
\begin{eqnarray}
J_{j} & \equiv & 
|p|^2 \lambda_j^2 + 2 {\rm Re}(pq^*) \lambda_j + |q|^2 
\nonumber \\
&=&  
C_{j} + 2 A^{(I)}_{j} \cos \delta + 2 A^{(II)}_{j} \cos 2 \delta  + 
2 B^{(I)}_{j} \sin \delta + 2 B^{(II)}_{j} \sin 2 \delta  
\label{Xj}
\end{eqnarray}
\begin{eqnarray}
C_{j}  &=&  
(p_{0}^2 + p_{1}^2) \lambda_j^2 + 
2 \{ p_{0} {\rm Re}(q_{0}) + p_{1} {\rm Re}(q_{1}) \} \lambda_j + 
|q_{0}|^2 + |q_{1}|^2 + |q_{2}|^2 
\nonumber \\
A^{(I)}_{j} &=& 
p_{0} p_{1} \lambda_j^2 + 
\{ p_{0} {\rm Re}(q_{1} + q_{2})  + p_{1} {\rm Re}(q_{0}) \} \lambda_j + 
{\rm Re}(q_{0}) {\rm Re}(q_{1} + q_{2}) + {\rm Im}(q_{0}) {\rm Im}(q_{1} + q_{2}) 
\nonumber \\
A^{(II)}_{j} &=& 
p_{1} {\rm Re}(q_{2}) \lambda_j + {\rm Re}(q_{1}) {\rm Re}(q_{2})  + {\rm Im}(q_{1}) {\rm Im}(q_{2})   
\nonumber \\
B^{(I)}_{j} &=& 
\{ p_{0} {\rm Im}(q_{1} - q_{2})  - p_{1} {\rm Im}(q_{0}) \} \lambda_j +  
{\rm Re}(q_{0}) {\rm Im}(q_{1} - q_{2}) - {\rm Im}(q_{0}) {\rm Re}(q_{1} - q_{2}) 
\nonumber \\
B^{(II)}_{j} &=& 
- p_{1} {\rm Im}(q_{2}) \lambda_j + {\rm Im}(q_{1}) {\rm Re}(q_{2}) - {\rm Re}(q_{1}) {\rm Im}(q_{2})  
\label{ABCI-II}
\end{eqnarray}
$\tilde{J}$ is given by 
\begin{eqnarray}
\tilde{J} = \frac{ 1 }{\Delta_{12} \Delta_{23} \Delta_{31}} 
\Bigl[
J^{(I)} \sin \delta + J^{(II)} \sin 2 \delta + 
K^{(0)} + K^{(I)} \cos \delta + K^{(II)} \cos 2 \delta 
\Bigr]
\end{eqnarray}
 where 
\begin{eqnarray}
J^{(I)} &=& \{ p_{0} {\rm Re}(q_{1} - q_{2})  - p_{1} {\rm Re}(q_{0}) \} 
\nonumber \\
J^{(II)} &=& - 
p_{1} {\rm Re}(q_{2}) 
\nonumber \\
K^{(0)} &=&  
- p_{0} {\rm Im} q_{0} - p_{1} {\rm Im} q_{1} 
\nonumber \\
K^{(I)} &=& 
-  \{ p_{0} {\rm Im}(q_{1} + q_{2})  + p_{1} {\rm Im}(q_{0}) \} 
\label{J}
\nonumber \\
K^{(II)} &=& -  p_{1} {\rm Im}(q_{2})  
\label{JK}
\end{eqnarray}

The coefficients $p$ and $q$, which are defined in (\ref{KTYeq2}), 
can be written as 
\begin{eqnarray} 
p &=& p_{0} + p_{1} e^{- i \delta} 
\nonumber \\
q &=& q_{0} + q_{1} e^{- i \delta} + q_{2} e^{+ i \delta} 
\label{pq}
\end{eqnarray}
where 
\begin{eqnarray}
p_{0} &=& \delta_{21} c_{12} s_{12} c_{23} c_{13}, 
\nonumber \\
p_{1} &=& ( \delta_{31} - s^2_{12} \delta_{21} ) s_{23} c_{13} s_{13}
\nonumber \\
q_{0} &=& - \delta_{31} \delta_{21} c_{12} s_{12} c_{23} c_{13} +  
\tilde{ \varepsilon}_{e \tau} c_{23} s_{23} 
\Bigl[ 
\delta_{31} c^2_{13}  -  \delta_{21} ( c^2_{12} - s^2_{12} s^2_{13} ) 
\Bigr].  
\nonumber \\
q_{1} &=& 
\delta_{21} \Bigl[ 
- \delta_{31} c^2_{12} s_{23} c_{13} s_{13} + 
\tilde{ \varepsilon}_{e \tau} c_{12} s_{12} s^2_{23} s_{13} 
\Bigr], 
\nonumber \\
q_{2} &=& - \tilde{ \varepsilon}_{e \tau} \delta_{21} c_{12} s_{12} c^2_{23} s_{13}. 
\label{p01q012}
\end{eqnarray}
Notice that the coefficient $p$ is identical with the standard case 
without NSI interactions whereas 
$q$ has a little more complex $\delta$ dependence with $q_{2}$ term  
and the coefficients $q_{i}$ ($i=1-3$) have imaginary parts.

Collecting formulae given in the equations from (\ref{Pemu_exact}) 
to (\ref{p01q012}) and using the exact eigenvalues by solving the 
cubic equation (\ref{eigenvalue-eq}), 
we obtain the exact expression of the appearance probability 
$P(\nu_{e} \rightarrow \nu_{\mu})$ with the neutrino NSI 
$ \varepsilon_{e e}$ and $ \varepsilon_{e \tau}$. 
Notice that $\tilde{a}$ and $\tilde{ \varepsilon}_{e \tau}$ 
are the renormalized quantities defined by (\ref{renormalized}). 
\begin{eqnarray}
&& P(\nu_e \to \nu_{\mu}; \varepsilon_{e e}, \varepsilon_{e \tau}) 
\nonumber \\
&=&  
8 \sum_{(ijk)}^{{\rm cyclic}} 
\frac{ - 1 }{ ( \Delta_{ij} \Delta_{jk} )^2  } 
\Bigl[
\frac{1}{2} C_{j} + A^{(I)}_{j} \cos \delta + A^{(II)}_{j} \cos 2 \delta  + 
B^{(I)}_{j} \sin \delta + B^{(II)}_{j} \sin 2 \delta 
\Bigr] 
\nonumber \\
&\times&  
\cos \left(\frac{ \tilde{a} L}{4E} \Delta_{ki} \right) 
\sin \left(\frac{ \tilde{a} L}{4E} \Delta_{ij} \right) 
\sin \left(\frac{ \tilde{a} L}{4E} \Delta_{jk} \right), 
\nonumber \\
&+& 
8 \frac{ 1 }{\Delta_{12} \Delta_{23} \Delta_{31}} 
\Bigl[
J^{(I)} \sin \delta + J^{(II)} \sin 2 \delta + 
K^{(0)} + K^{(I)} \cos \delta + K^{(II)} \cos 2 \delta 
\Bigr] 
\nonumber \\
&\times&
\sin \left(\frac{ \tilde{a} L}{4E} \Delta_{12} \right) 
\sin \left(\frac{ \tilde{a} L}{4E} \Delta_{23} \right) 
\sin \left(\frac{ \tilde{a} L}{4E} \Delta_{31} \right). 
\label{Pemu_exact2}
\end{eqnarray}
%

\subsection{Leading order formula of $P(\nu_e \to \nu_{\mu})$} 

Now, we define the perturbative scheme we use in our derivation 
of the approximate expression of the appearance probability 
$P(\nu_{e} \rightarrow \nu_{\mu})$. 
We regard $\delta_{31}$ and $a$ as of order unity, and assume that 
$s_{13} \simeq  \delta_{21} \simeq  \tilde{ \varepsilon}_{e \tau} $ 
are small and are of the same order to organize the perturbation theory. 
We denote the expansion parameters symbolically as $\epsilon$. 
We assume that $\epsilon \sim 10^{-2}$. 
If we organize the perturbation expansion in terms of $\epsilon$ 
we will recognize that the leading order terms in the oscillation 
probability in $P(\nu_e \to \nu_{\mu})$ is of order $\epsilon^2$.

To the next to leading order in $\epsilon$ 
the solutions of the equation are given under the convention 
that $\lambda_1 < \lambda_2 < \lambda_3$ by 
\begin{eqnarray}
\lambda_1 &=& c_{12}^2 \delta_{21} ,
\nonumber \\
\lambda_2 &=& \delta_{31} - 
\frac{\delta_{31}}{1 - \delta_{31} } 
\Bigl\{
s_{13}^2 + 2 c_{23} s_{13} {\rm Re}( \tilde{ \varepsilon}_{e \tau} e^{i \delta} ) 
\Bigr\}
\nonumber \\
\lambda_3 &=& 1 + 
\frac{\delta_{31}}{1 - \delta_{31} } 
\Bigl\{
s_{13}^2 + 2 c_{23} s_{13} {\rm Re}( \tilde{ \varepsilon}_{e \tau} e^{i \delta} ) 
\Bigr\}   + 
s_{12}^2 \delta_{21} 
\label{eigenvalue}
\end{eqnarray}
In (\ref{eigenvalue}) we have ignored even smaller corrections to 
the lowest eigenvalue because it is of order $\epsilon^3$.

We restrict ourselves to the leading order as was done 
by Cervera {\it et al.} \cite{golden} for the standard case without 
$\varepsilon_{\alpha \beta}$'s. 
To this order, we in fact do not need the order $\epsilon$ corrections 
in (\ref{eigenvalue}). 
By putting back the physical quantities replacing the scaled variables,  
the $\nu_{\mu}$ appearance probability can be expressed to 
order $\epsilon^2$ as 
\begin{eqnarray}
&& P(\nu_e \to \nu_{\mu}; \varepsilon_{e e}, \varepsilon_{e \tau}) \vert_{2nd}=  
P(\nu_e \to \nu_{\mu}; \varepsilon = 0) \vert_{2nd} 
\nonumber \\
& & - \frac{4 c_{23} s_{23}^2}{(\tilde{a} - \Delta m_{31}^2)^2}
  \Bigl[ 2 \tilde{a} \Delta m_{31}^2 s_{13} \text{Re}(\varepsilon_{e\tau} e^{i\delta})
  + c_{23} \tilde{a}^2 |\varepsilon_{e\tau}|^2 \Bigr] 
\nonumber \\
&& \hspace{5.0cm} \times 
 \cos\left( \frac{\tilde{a} L}{4E} \right) \sin\left( \frac{\Delta m_{31}^2 L}{4E} \right)
  \sin\left( \frac{L}{4E}(\tilde{a} - \Delta m_{31}^2) \right) 
  \nonumber \\
 & & + 4 c_{23} s_{23}
  \left[ \frac{(\Delta m_{31}^2)^2}{(\tilde{a} - \Delta m_{31}^2)^2} s_{23}
   \left(  2 s_{13} \text{Re}(\varepsilon _{e\tau} e^{i\delta})
  + c_{23} |\varepsilon_{e\tau}|^2 \right)
 + \frac{2  \Delta m_{31}^2\Delta m_{21}^2 }{\tilde{a} (\tilde{a} - \Delta m_{31}^2)} c_{12}  s_{12} c_{23} \text{Re}(\varepsilon_{e\tau}) \right] 
 \nonumber \\
&& \hspace{5.0cm} \times 
 \cos\left( \frac{\Delta m_{31}^2 L}{4E} \right)  
 \sin\left( \frac{\tilde{a} L}{4E} \right) 
 \sin\left( \frac{L}{4E}(\tilde{a} - \Delta m_{31}^2) \right)
  \nonumber \\
 && + 4 c_{23} s_{23} \left[ c_{23} s_{23} |\varepsilon_{e \tau}|^2 - 2 \frac{\Delta m_{21}^2}{\tilde{a}} c_{12} s_{12} c_{23}  \text{Re}(\varepsilon_{e\tau}) \right] 
 \nonumber \\
&& \hspace{5.0cm} \times  
\cos\left( \frac{L}{4E}(\tilde{a} - \Delta m_{31}^2) \right) 
\sin\left( \frac{\tilde{a} L}{4E} \right)
\sin\left( \frac{\Delta m_{31}^2 L}{4E} \right) 
  \nonumber \\
 && - \frac{8 c_{23} s_{23}}{(\tilde{a} - \Delta m_{31}^2)}
  \Bigl[ \Delta m_{31}^2 s_{23} s_{13} \text{Im}(\varepsilon_{e \tau} e^{i\delta}) + \Delta m_{21}^2 c_{12} s_{12} c_{23}  \text{Im}(\varepsilon_{e \tau})  \Bigr] 
  \nonumber \\
&& \hspace{5.0cm} \times  
\sin\left( \frac{\Delta m_{31}^2 L}{4E} \right) 
\sin\left( \frac{\tilde{a} L}{4E} \right) 
\sin\left( \frac{L}{4E}(\tilde{a} - \Delta m_{31}^2) \right), 
   \label{Penu-2nd_etau}
\end{eqnarray}
where $P(\nu_e \to \nu_{\mu}; \varepsilon = 0) \vert_{2nd} $ is nothing but 
the Cervera {\it et al.} formula \cite{golden}
\begin{eqnarray}
&&  
P(\nu_e \to \nu_{\mu}; \varepsilon = 0) \vert_{2nd} = 
4 \frac{ ( \Delta m^2_{31} )^2 }{  (  \tilde{a} - \Delta m^2_{31} )^2  } 
s^2_{23} s^2_{13} \sin^2 \left(\frac{ L}{4E} ( \tilde{a} - \Delta m^2_{31})  \right)  
\nonumber \\
&& + 8 J_r \frac { \Delta m^2_{31}  \Delta m^2_{21}  }{  \tilde{a} ( \tilde{a} - \Delta m^2_{31} )} 
\sin \left(\frac{  \tilde{a} L}{4E}  \right) 
\sin \left(\frac{ L}{4E} ( \tilde{a}  - \Delta m^2_{31})  \right)  
\cos \left(  \delta -  \frac{  \Delta m^2_{31} L}{4E}   \right) 
\nonumber \\
&& + 
4 \left( \frac{ \Delta m^2_{21} } { \tilde{a} } \right)^2 
c^2_{12} s^2_{12} c^2_{23} 
\sin^2 \left(\frac{ \tilde{a} L}{4E}  \right).  
\label{Penu-zeroeps}
\end{eqnarray}
It is notable that the NSI effects survive in the leading order, 
$\simeq \epsilon^{2}$. 
If one want to have explicit expression with $\varepsilon_{e \tau}$ 
and $\varepsilon_{e e}$ one can just use the relations (\ref{renormalized}) in 
(\ref{Penu-2nd_etau}). 
The formula for $P(\nu_e \to \nu_{\mu})$ is valid only for small 
$\varepsilon_{e \tau}$ but for any finite size $\varepsilon_{e e}$. 
The antineutrino probability can be obtained by the replacement 
$\delta \rightarrow - \delta$, 
$a \rightarrow - a $, and 
$\varepsilon_{\alpha \beta} \rightarrow \varepsilon_{\alpha \beta}^{*}$. 
We have checked that the same analytic formulas are obtained, 
when expressed in terms of observable physical quantities,   
even if we work in the intermediate energy region where 
$\lambda_2 > \lambda_3$.

Similarly, the formula of $P(\nu_e \to \nu_{\mu})$ with 
$\tilde{ \varepsilon}_{e \mu}$ can be computed to the leading order as 
\begin{eqnarray}
&& P(\nu_e \to \nu_{\mu}; \varepsilon_{e e}, \varepsilon_{e \mu}) \vert_{2nd}= 
P(\nu_e \to \nu_{\mu}; \varepsilon = 0) \vert_{2nd} 
\nonumber \\ 
&& - \frac{ 4 \tilde{a} s_{23}^3}{(\tilde{a} - \Delta m_{31}^2)^2}
 \Bigl[  2 \Delta m_{31}^2 s_{13} \text{Re}(\varepsilon_{e\mu} e^{i\delta})
  + s_{23} \tilde{a} |\varepsilon_{e\mu}|^2 \Bigr] 
  \nonumber \\ 
&& \hspace{5.0cm} \times   
\cos\left( \frac{\tilde{a} L}{4E} \right) \sin\left( \frac{\Delta m_{31}^2 L}{4E} \right)
  \sin\left( \frac{L}{4E}(\tilde{a} - \Delta m_{31}^2) \right) 
  \nonumber \\ 
   && + 4 \frac{   (\tilde{a} - c_{23}^2 \Delta m_{31}^2)  }{(\tilde{a} - \Delta m_{31}^2)^2}
  \left[ 2 \Delta m_{31}^2 s_{23} s_{13} \text{Re}(\varepsilon_{e\mu} e^{i\delta})
  + (\tilde{a} - c_{23}^2 \Delta m_{31}^2) |\varepsilon_{e\mu}|^2 \right. 
  \nonumber \\ 
  && \left. 
   \hspace{6.4cm} 
   + 2 (\tilde{a} - \Delta m_{31}^2) 
   \left( \frac{ \Delta m_{21}^2 } {\tilde{a} } \right) 
   c_{12} s_{12} c_{23} \text{Re}(\varepsilon_{e\mu}) \right] 
  \nonumber \\ 
&& \hspace{5.0cm} \times   
\cos\left( \frac{\Delta m_{31}^2 L}{4E} \right) 
 \sin\left( \frac{\tilde{a} L}{4E} \right) 
 \sin\left( \frac{L}{4E}(\tilde{a} - \Delta m_{31}^2) \right)
  \nonumber \\ 
  && + 4 c_{23}^3
  \left[ c_{23} |\varepsilon_{e\mu}|^2
  + 2 \frac{\Delta m_{21}^2}{\tilde{a}} c_{12} s_{12} \text{Re}(\varepsilon_{e\mu}) \right] 
  \nonumber \\ 
&& \hspace{5.0cm} \times  
 \cos\left( \frac{L}{4E}(\tilde{a} - \Delta m_{31}^2) \right) 
  \sin\left( \frac{\tilde{a} L}{4E} \right)
  \sin\left( \frac{L}{4E} \Delta m_{31}^2 \right) 
  \nonumber \\ 
 && + \frac{8 c_{23} s_{23}}{(\tilde{a} - \Delta m_{31}^2)}
 \Bigl[  \Delta m_{31}^2 c_{23} s_{13} \text{Im}(\varepsilon_{e \mu} e^{i\delta})
   - \Delta m_{21}^2 c_{12} s_{12} s_{23} \text{Im}(\varepsilon_{e \mu}) \Bigr]  
   \nonumber \\ 
&& \hspace{5.0cm} \times  
 \sin\left( \frac{\Delta m_{31}^2 L}{4E} \right) 
  \sin\left( \frac{\tilde{a} L}{4E} \right)
 \sin\left( \frac{L}{4E}(\tilde{a} - \Delta m_{31}^2) \right). 
\label{Penu-2nd_emu}
\end{eqnarray}
%



\end{document}